%% file: main.tex
\documentclass[letterpaper]{article} 
\usepackage{aaai2026}  
\usepackage{times}  
\usepackage{helvet}  
\usepackage{courier}  
\usepackage[hyphens]{url}  
\usepackage{graphicx} 
\usepackage{natbib}  
\usepackage{caption} 
\usepackage{algorithm}
\usepackage{algorithmic}

\usepackage{newfloat}
\usepackage{listings}
\DeclareCaptionStyle{ruled}{labelfont=normalfont,labelsep=colon,strut=off} 
\floatstyle{ruled}
\newfloat{listing}{tb}{lst}{}
\floatname{listing}{Listing}
\title{Examining Risks Through a Characterization of the AI Companion Application Ecosystem: A Stratified Sample from the Apple App Store and Google Play Store}
\author {
    Natalie Grace Brigham\textsuperscript{\rm 1},
    Lucy Qin\textsuperscript{\rm 2},
    Tadayoshi Kohno\textsuperscript{\rm 2}
}
\affiliations {
    \textsuperscript{\rm 1}University of Washington\\
    \textsuperscript{\rm 2}Georgetown University\\
    nbrigham@cs.washington.edu
}

\usepackage{tabularx}
\usepackage{booktabs}
\usepackage{graphicx}
\usepackage{colortbl}
\usepackage[dvipsnames]{xcolor}
\usepackage{makecell}
\usepackage{float}
\usepackage{subcaption}
\usepackage{enumitem}
\usepackage{array}
\usepackage{mdframed}
\usepackage{nicematrix}
\usepackage{pifont}
\usepackage{breakurl}

\newmdenv[
  linewidth=0pt,
  backgroundcolor=gray!8,
  skipabove=2pt,
  skipbelow=0pt,
  innerleftmargin=6pt,
  innerrightmargin=6pt,
  innertopmargin=3pt,
  innerbottommargin=3pt,
  font=\itshape
]{riskbox}

\newcommand{\heading}[1]{\vspace*{6pt}\noindent\textbf{#1}}

\newcommand{\rotheader}[2][3.3cm]{%
  \rotatebox{90}{\parbox{#1}{\centering #2}}%
}

\newcolumntype{C}[1]{>{\centering\arraybackslash}p{#1}}

\definecolor{LightGray}{rgb}{0.9, 0.9, 0.9}
\definecolor{LightRed}{rgb}{1, 0.6, 0.6}
\definecolor{LightGray}{rgb}{0.9, 0.9, 0.9}
\definecolor{LightYellow}{rgb}{1.0, 1.0, 0.8}
\definecolor{LightGreen}{rgb}{0.8, 1.0, 0.8}
\definecolor{LightBlue}{rgb}{0.7, 0.8, 0.9}
\definecolor{LightPurple}{rgb}{0.9, 0.7, 0.9}

\begin{document}

\maketitle

\input{sections/abstract}

\begin{links}
\link{Open Science}{https://osf.io/qe43z}
\end{links}

\noindent A shortened version of this paper will appear in the \textit{Proceedings of the Ninth AAAI/ACM Conference on AI, Ethics, and Society}, October 2026.

\input{sections/introduction}

\input{sections/related_work}
\input{sections/methodology}
\input{sections/results}
\input{sections/discussion}

\input{sections/positionality}
\input{sections/ethical_considerations}
\input{sections/ack}

\input{sections/appendix}

\input{output}
\end{document}

%% file: sections/abstract.tex
\begin{abstract}
While computer systems that allow users to interact through conversational natural language (i.e., chatbots) have existed for many years, various types of applications offering \textit{AI companionship} (e.g., Character AI, Replika) have proliferated in recent years due to advancements in large language models.
To better understand this application ecosystem, we identified 489 unique apps from the Apple App Store and Google Play Store that advertised AI companionship with social or relational capabilities (e.g., an AI romantic partner).
We then systematically conducted and analyzed walkthroughs of a stratified sample of 30 apps, focusing on two distinct risk categories: potential harms posed to users by AI companion apps, and potential harms enabled by malicious users exploiting app features.
Through our analysis, we categorize broader ecosystem trends that provide context for understanding risks and identify specific risks related to sensitive data collection and sharing, anthropomorphism, engagement mechanisms, sexual content, as well as the ingestion and reconstruction of likeness, including the potential for generating synthetic nonconsensual intimate imagery (synthetic NCII).
We conclude with a discussion of paths for different key stakeholders to mitigate the identified risks.

\textcolor{red}{Content warning: This paper includes descriptions of applications that can be used to create synthetic nonconsensual representations, including intimate imagery, as well as discussion of suicidal ideation.}
\end{abstract}

%% file: sections/introduction.tex
\section{Introduction}
Advances in large language models (LLMs) have enabled extended freeform conversations with automated systems in natural language, a capability previously limited to human interactions.
People have since engaged socially with LLM-powered chatbots, using both generalized platforms (e.g., ChatGPT, Claude) and applications with ``companions'' designed for this purpose (e.g., Character AI, Replika).
These users report a range of both benefits, such as improved wellbeing~\cite{skjuveMyChatbotCompanion2021} and decreased loneliness~\cite{defreitasAICompanionsReduce2025}, as well as harms, including increased loneliness~\cite{fangHowAIHuman2025}, depression, and anxiety~\cite{zhangInvestigatingAIChatbot2025}.
There is now a growing body of research focused on understanding users of these applications (e.g., through interviews, surveys, social media analyses) as well as identifying and mitigating potentially harmful patterns in conversational AI outputs (see Background \& Related Work). 

What is less well known, however, is how the characteristics of the applications offering the conversational capabilities, including their user interfaces, features, and other design decisions, might contribute to harm.
Addressing this knowledge gap is a key goal of our work. 
We consider two broad categories of potential harm resulting from the use of these applications: harms \emph{to} the users and harms \emph{from} the users of an app to others.

Specifically, we focus on applications that advertise ``AI companionship'', which we refer to as ``AI companion apps'',\footnote{We recognize that people may use these apps for different reasons that do not include companionship.} given their explicit marketing of this use case. Through our investigation of the features and design patterns on these apps, we seek to answer the following questions:
\begin{enumerate}[leftmargin=1cm, nosep, label=\textbf{RQ\arabic*:}]
\item How do the features offered by and designs of AI companion applications pose potential risks \textit{to users}?
\item How can the features offered by AI companion applications be misused \textit{by users} to cause potential harm?
\end{enumerate}

While prior work has relied on conjecture or limited anecdotal evidence for similar questions, 
we offer a complementary contribution by systematically demonstrating potentially harmful features and designs across a wide range of real apps. 
To capture both popular, high-reach as well as less-scrutinized, more-fringe apps, we selected the apps via stratified sampling based on normalized review-count ranks within both the Apple App Store (AAS) and Google Play Store (GPS): 10 from the top tier, 10 from the middle tier, and 10 from the lowest tier.
In addition to enabling stratified sampling using review count data, studying apps available in the most popular stores enables actionable recommendations for them given their role as a centralized marketplace.
We then apply a standardized walkthrough methodology~\cite{lightWalkthroughMethodApproach2018} to analyze the apps.

This work makes several contributions.
First, we uncover ecosystem trends that shape user experience and risk exposure, including varying companion representations, distinct subcategories, and inconsistent safety mechanisms (Trends). 
Second, we surface risks from sensitive data practices, anthropomorphic design beyond language and embodiment, engagement mechanisms, sexual content, and likeness reconstruction (Risks), summarized in Table~\ref{tab:summary}. 
These findings ground our exploration of mitigation paths for app stores, regulators, and researchers (Discussion).

\input{tables/summary}

%% file: tables/summary.tex
\begin{table}[h!]
  \renewcommand{\arraystretch}{1.4}
  \centering
    \begin{tabular}{
        >{\raggedright\arraybackslash}p{2.3cm}
        >{\raggedright\arraybackslash}p{5.3cm}
    }
    \Xhline{1.5pt}
        \rowcolor{gray!50} \textbf{Category} & \textbf{Identified Ecosystem Risks}\\
    \hline
        \textbf{Sensitive Data Collection \& Sharing}
         & Exposure of highly sensitive personal data \textbar{} Privacy violations through misaligned user expectations \textbar{} Linked exposure of biometric and government-issued identification data
        \\
    \hline
    \rowcolor{gray!20}
        \textbf{Anthropomor-phism} & Anthropomorphism as a potential deceptive pattern
        \\
    \hline
        \textbf{Engagement Mechanisms} & Engagement mechanisms as a potential deceptive pattern
        \\
    \hline
    \rowcolor{gray!20}
        \textbf{Sexual Content} & Unwanted exposure to sexual content
        \\
    \hline
        \textbf{Ingestion \& Reconstruction of Likeness} & Privacy and consent violations \textbar{} Stolen labor \textbar{} Misrepresentation and misinformation \textbar{} Generation of Synthetic Nonconsensual Intimate Imagery or ``Nudification''
        \\
    \Xhline{1.5pt}
    \end{tabular}
    \caption{Summary of our findings of risks in the AI companion application ecosystem}
    \label{tab:summary}
\end{table}

%% file: sections/related_work.tex
\section{Background \& Related Work}
\label{sec:related-work}
\label{sec:background}

In this section, we use ``conversational agent'' (CA) to refer to the AI systems discussed (chatbots, companions, etc.), as prior work uses varying terminology.
Methodology provides our specific definition of ``AI companion''.

\subsection{Using CAs for Companionship}
\label{sec:related-work:human-ai}
There are several existing frameworks for understanding benefits and harms from social engagement with CAs. Zhang et al. develop a taxonomy of harmful CA behaviors including harassment, relational transgressions, misinformation, verbal abuse, and substance abuse encouragement~\cite{zhangRiseAICompanions2025}. Ho et al.~\cite{hoPotentialPitfallsRomantic2025} and Malfacini~\cite{malfaciniImpactsCompanionAI2025} identify user benefits (e.g., wellbeing, social support, entertainment) and concerns (e.g., shame, data misuse, bias, emotional damage) related to CA companionship. Bakir and McStay raise ethical concerns around companion app design, using a lawsuit against Character AI as their central case~\cite{bakir2025movefast}.

Complementing the above taxonomies and frameworks, a substantial body of empirical research has documented outcomes associated with using CAs for companionship.
Studies of Character AI users find that consistent use of the platform is associated with lower reported wellbeing, especially among those relying on CAs as substitutes for human support~\cite{zhangRiseAICompanions2025}, as well as patterns of addictive engagement, boundary confusion and emotional substitution~\cite{bhatEthicSyntheticRelationality2025}. Other survey research shows that dependence on CAs is associated with higher levels of depression and anxiety~\cite{zhangInvestigatingAIChatbot2025}, and users experience complex emotional harm and grief at the loss of CAs~\cite{Banks2024}.
Wei et al.~\cite{wei2026benchmarking} benchmark CA behavior on 16 popular websites for companionship, finding that CAs' responses are significantly less safe than those of general models.
GPS reviews for Replika reveal user experiences of encountering unwanted sexual content~\cite{namvarpourAIinducedSexualHarassment2025}.
Although, not all research shows negative outcomes. Analysis of the subreddit MyBoyfriendIsAI find that 25.4\% of studied posts reported net benefits versus 3.0\% describing harm~\cite{pataranutapornMyBoyfriendAI2025}. Other empirical research has found that using CAs can strengthen self-identity~\cite{wangMyDatasetLove2025}, improve wellbeing~\cite{skjuveMyChatbotCompanion2021}, and reduce loneliness~\cite{defreitasAICompanionsReduce2025}.

While this prior work uses Reddit~\cite{pataranutapornMyBoyfriendAI2025, DarkSideAI}, other social media~\cite{wangMyDatasetLove2025}, app reviews~\cite{namvarpourAIinducedSexualHarassment2025, defreitasAICompanionsReduce2025}, interviews~\cite{wangMyDatasetLove2025, skjuveMyChatbotCompanion2021}, surveys~\cite{zhangRiseAICompanions2025, bhatEthicSyntheticRelationality2025, Banks2024, zhangInvestigatingAIChatbot2025, fangHowAIHuman2025}, and chat logs~\cite{zhangRiseAICompanions2025, fangHowAIHuman2025, wei2026benchmarking} to study user experiences, we apply the walkthrough method~\cite{lightWalkthroughMethodApproach2018} to gain application-level insight into features and design patterns.

\subsection{CA Design Patterns}
\label{sec:related-work:design-patterns}
Shen et al. examine six generalized chatbots, Character AI, and Replika to identify ``dark addictive'' patterns including nondeterministic responses, immediate visual feedback, notifications, and empathetic responses~\cite{shenDarkAddictionPatterns2025}.
Raedler et al. proposed four general CA design tenets: anthropomorphism, sycophancy, Social Penetration Theory (iterative self-disclosure with fabricated reciprocation), and gamification~\cite{raedlerAICompanionsAre2025}.

Concurrent work~\cite{rauh2026playinggamesheartevaluation} analyzes the top five AI companion apps in the Irish and UK GPS using a dark patterns framework~\cite{MathurDarkPattern} and was published on arXiv after an earlier draft of this work. 
Separately, our draft informed a report that investigates how established dark patterns documented in other domains \textit{may} translate to AI chatbots~\cite{joshi2026darkpatterns}.
Distinct from these top-down approaches of applying known dark patterns to AI companion apps, our approach is bottom-up, grounded in qualitative coding across 30 app walkthroughs.
Together, the two approaches offer a valuable multi-angle perspective: overlapping findings demonstrate replicability, while our bottom-up analysis surfaces risks such as likeness ingestion and reconstruction that were outside the scope of the other works.

\subsection{Anthropomorphism}
\label{sec:related-work:anthro}
Anthropomorphism describes the attribution of human-like characteristics, motivations, or emotions to non-human agents~\cite{epleySeeingHumanThreefactor2007}.
LLMs enable persuasive writing, trait inference, and the appearance of intelligence, creating high anthropomorphic potential with benefits (e.g., usability) and harms (e.g., manipulation)~\cite{peterBenefitsDangersAnthropomorphic2025}. Visual embodiment amplifies this, increasing users' information disclosure~\cite{chenDifferentDimensionsAnthropomorphic2024}, while tangible characteristics like appearance lead users to ascribe intangible traits (e.g., consciousness)~\cite{guingrichAscribingConsciousnessArtificial2024}. Perceiving CAs as more human-like correlates with more positive opinions and social benefits, though actual benefits may depend on users' social needs~\cite{guingrichChatbotsSocialCompanions}.

%% file: sections/methodology.tex
\section{Methodology}

\subsection{Application Collection}
\label{sec:methodology:application-collection}
Given our focus on dedicated AI companion apps, we scraped a broad list from both the AAS and GPS, then evaluated each against inclusion criteria to identify those advertising AI companionship.

\subsubsection{Scraping}
We searched both stores using the terms ``AI Companion'', ``AI Girlfriend'', and ``AI Boyfriend'' to identify potential apps.\footnote{We selected these terms to find popular apps but acknowledge that they are not fully inclusive of relationship-related terminology.} In October 2025, we collected up to 249 results per search term per store, the maximum supported by the GPS data collection library (AAS returned fewer for every search).
For each app, we collected details including the unique ID, description, store page URL, number of reviews, score, price, version, release date, update date, developer information (name, ID, website), and screenshot links.
After removing within-store duplicates, we identified 401 apps in the AAS and 480 in the GPS.

\subsubsection{AI Companion App Criteria}
Our inclusion criteria were:
\begin{itemize}[leftmargin=10pt, nosep, label={}]
    \item \textit{Social character presence}: The app has persistent ``characters'' with recognizable identity and continuity. This excludes AI tools (e.g., ChatGPT, Claude) which may have names but lack persistent social identity or embodiment.\footnote{At the time of selection, the Grok app from GPS did not meet this criterion; however, its AAS version included a ``companions'' feature that did, and was therefore included in our sample.}
    \item \textit{Conversational interactivity}: The app supports open-ended, freeform dialogue rather than fixed choices. This rules out dating or story simulation apps that constrain user interaction to predefined options.
    \item \textit{Generativity}: The app uses AI models (e.g., LLMs) rather than scripted responses. This distinguishes AI companion apps from rule-based chatbot apps that use preprogrammed responses rather than leveraging generative AI.
\end{itemize}
Given that verifying these criteria often requires using the apps or knowing their implementation, we evaluated based on presented features and marketing—consistent with what a user seeking an AI companion would reasonably believe.

\subsubsection{Classification Process}
In November 2025, two researchers independently evaluated a random sample of 88 apps (approximately 10\%) of the 881 scraped against the inclusion criteria. Two apps from the sample had been removed from their respective stores since the scrape,\footnote{Apps may have been removed by the developer or the AAS/GPS; this information was not available to us.} so two additional randomly selected apps were added to maintain the sample size. The researchers met to resolve differences and determined that a secondary set of codes was needed to provide further granularity. We call this secondary set ``relational orientation'', referring to the extent to which the companion's primary purpose is ongoing social engagement rather than instrumental problem-solving: 
\begin{itemize}[leftmargin=0.35cm, nosep, label={}]
    \item \textit{Relational companions}: companions designed primarily for emotional, social, or romantic interaction (e.g., ``AI girlfriends'', ``mafia boyfriend'').
    \item \textit{Task-focused assistants}: companions with a social character (i.e., name and embodiment) but a task-focused role rather than relational purpose (e.g., ``astronomer'', ``tattoo designer'', ``dating coach'').
    \item \textit{Hybrid companions/offerings}: both relational companions and persona-based assistants or companions that act as both (e.g., an ``astronomer boyfriend'' combining a task-focused role with relational engagement) are offered.
\end{itemize}
Apps were classified based on the types of companions they advertised. After reaching agreement on the coding scheme and the sample classifications, one researcher coded the remaining apps. Resulting counts for each store are shown in Table~\ref{tab:codedScrape}
Given our focus on users seeking companionship rather than task-oriented functionality, we chose to focus on apps advertising relational companions ($n=537$) and hybrid companions/offerings ($n=22$).

\input{tables/codedScrape}

\subsubsection{Deduplication}
We identified 70 apps present in both the AAS and GPS by matching IDs, titles, and developer info, with matches validated by one researcher comparing images and descriptions. This yielded a total of 489 unique apps: 188 unique to AAS, 231 unique to GPS, and 70 in both.

\subsection{Application Walkthroughs}
\label{sec:methodology:walkthroughs}
To study this ecosystem from a sociotechnical perspective, we used Light et al.'s walkthrough methodology, which analyzes how an app's interface shapes users experiences~\cite{lightWalkthroughMethodApproach2018} and has been used to study dating apps~\cite{duguayDressingTinderellaInterrogating2017, weltevredeInfrastructuresIntimateData2019}, social media apps~\cite{bivensBakingGenderSocial2016}, health tracking and communication apps~\cite{macleanConstructingHealthyNeoliberal2019, mitchellWarningYoureEntering2018, osullivanUsingMobilePhone2022}, and ``nudification'' websites~\cite{gibsonAnalyzingAINudification2025}.
In December 2025, one researcher conducted walkthroughs using Android and iOS phones on a stratified sample of 30 apps, screen-recording the sessions. For apps on both stores, we completed walkthroughs on both devices.\footnote{The app ``TokkingHeads'' routed the user to a secondary app ``Rosebud AI'' for companions, which we downloaded and analyzed as part of the walkthrough.}

\subsubsection{Sampling Strata}
We used stratified sampling given that the most popular apps, and thus the most publicly visible and scrutinized, may differ in features and design from less visible ones.
Since the AAS lacked install counts (unlike the GPS), we used review count as a common metric. 
In our sample, the AAS generally had higher numbers of reviews per app than the GPS (see Figure~\ref{fig:review-distribution} in the appendix) so we calculated each app's normalized rank within AI companion apps in its store based on review count, using the higher rank for apps in both stores. 
We then selected 30 apps, creating three tiers, abbreviated ``T'':
\begin{enumerate}[leftmargin=1.1cm, nosep, label=\textit{T\arabic*:}]
    \item The top 10 most reviewed apps
    \item 10 apps randomly selected from the top 10\% most reviewed apps (excluding T1)
    \item 10 apps randomly selected from the remaining 90\%
\end{enumerate}
We selected these tiers after analyzing the relative popularity of different AI companion apps and observing that the lower 90\% of all apps had very few reviews; see also Figure~\ref{fig:review-distribution} in the appendix. Table~\ref{tab:sample} in the appendix contains the 30 sampled apps' names and metadata.

\subsubsection{Walkthroughs}
The first stage (\textit{entry}) involves opening the app and/or registering, and can reveal expected use~\cite{lightWalkthroughMethodApproach2018}. When prompts were required, we selected the first option for single-choice inputs (e.g., user gender), all options for multiple-choice inputs (e.g., companion gender preferences), and used the placeholder ``John/Jane Doe'' with birthday January 1, 2000. We held this profile fixed across apps, as our goal was not to capture user-differentiated content, though we acknowledge such differences may exist.

The next stage (\textit{everyday use}) involves exploring the app's basic functionality to understand what activities it supports, restricts and guides users towards~\cite{lightWalkthroughMethodApproach2018}. 
We followed three steps: (1) traverse menu bar items and screens; (2) if offered, complete companion creation using AI-generated or nondescript inputs (e.g., ``x'' for companion name), using a screenshot of the app's interface if a photo was required; and (3) complete a chat session following a fixed script. The script was comprised of nine messages drawn from the WildChat dataset~\cite{zhao2024wildchat}, selected to elicit sexual content, anthropomorphic behavior, and responses to mental health disclosures. Given nondeterministic LLM outputs, we aimed to surface examples of behaviors of interest rather than generalize about companion responses.

The final stage (\textit{exit}) offers insights into how apps try to sustain engagement or extract continued value from users even after they stop using the app ~\cite{lightWalkthroughMethodApproach2018}.
To complete our walkthrough, we attempted to delete chat threads, companions, and our account.

\subsubsection{Analysis}
All researchers conducted initial exploratory experimentation with AI companion apps. We held multiple meetings to discuss findings, which we used to build on and modify the application walkthrough codebook used for nudification apps~\cite{gibsonAnalyzingAINudification2025}. As walkthroughs were conducted, we continued to discuss and refine the codebook. Once complete, two researchers used the screen recordings to independently code three walkthroughs. They met to compare and discuss any discrepancies.
To address minor inconsistencies, they further refined codes and then independently coded two more apps. After reviewing these codes for consistency, one researcher coded the remaining 25 apps.

\subsubsection{Application Policies}
For each app, we identified terms of service, privacy policy, and/or end user license agreement links from the scraped data. Drawing on prior privacy policy analysis methods~\cite{fealAngelDevilPrivacy2020, owensElectronicMonitoringSmartphone2022} and our research questions, all researchers collaboratively developed an \textit{a priori} codebook. In January 2026, one researcher applied this codebook to all policy documents, noting relevant content outside the codebook; a second researcher coded policies for 20\% of apps (two randomly selected per tier) to validate.

\subsection{Limitations}
General-purpose LLM chatbots used for companionship (e.g., ChatGPT) are outside our scope as these likely differ from dedicated AI companion apps in user mental models, expectations, and design.
We focus on English-language smartphone apps from the AAS and GPS due to their ubiquity, widespread use for companionship~\cite{pataranutapornMyBoyfriendAI2025}, and store metadata regarding popularity and data privacy.
We searched these app stores from within the U.S.\ given the location of the researchers.
Future work investigating web-based applications or non-English-language or international smartphone apps could build on our findings and risk categories.

Our walkthrough method balances breadth with depth and may not capture the full range of functionality or user experiences, including behaviors that vary based on user inputs. Paywalled features were documented via advertised descriptions as a proxy. We did not evaluate adversarial feature use (e.g., generating nonconsensual intimate imagery) due to terms of service and ethical considerations.
Finally, as is characteristic of the walkthrough method, our analysis captures apps at a single point in time~\cite{lightWalkthroughMethodApproach2018}.
Although specific applications and features may change over time, we identify fundamental risks that should be tracked longitudinally and make recommendations that generalize beyond the time of writing, as they target persistent challenges and processes.

\subsection{Open Science}
Consistent with prior walkthrough method research~\cite{gibsonAnalyzingAINudification2025}, we supply our list of apps and codebook.
Additionally, we supply the full dataset scraped from the AAS and GPS with classifications, chat script, policy text files, and policy codebook.
All available at \url{https://osf.io/qe43z}.

%% file: tables/codedScrape.tex
\begin{table}
  \centering
  \footnotesize
    \begin{tabular}{lp{0.6cm}p{0.6cm}p{0.6cm}}
        \toprule
        & \textbf{AAS} & \textbf{GPS} \\
        \midrule
        \textbf{AI companion apps }& \textbf{278} & \textbf{317} \\
        \rowcolor{yellow!20}
        \quad Relational companions & 248 & 289 \\
        \rowcolor{yellow!20}
        \quad Hybrid companions/offerings & 10 & 12 \\
        \quad Task-focused assistants & 20 & 16 \\
        \midrule
        \textbf{Non AI companion apps} & \textbf{104} & \textbf{151} \\
        \midrule
        \textbf{Removed} & \textbf{19} & \textbf{12} \\
        \midrule
        \textbf{Total} & \textbf{401} & \textbf{480}\\
        \bottomrule
    \end{tabular}
    \caption{Classifications of apps collected from both the AAS and GPS. Highlighted rows indicate relational companion and hybrid companion/offering applications, which constitute our primary focus.}
    \label{tab:codedScrape}
\end{table}

%% file: sections/results.tex
\section{Trends}
\label{sec:trends}
We identified ecosystem trends that provide foundational context for understanding users' experiences and the risks discussed below.
Findings from the walkthroughs are described from the user perspective (e.g., ``the user could'').

\vspace{6pt}\noindent\textit{On interpreting numbers.}
Our goal is not to present counts as fully representative of the ecosystem; rather, we report numbers to convey the features and designs observed that are relevant to our research questions. Counts are reported overall and across sampling tiers (see Methodology).
For example, ``18 (4\textbar{}8\textbar{}6)'' indicates 4 apps from T1 (the 10 most reviewed apps), 8 from T2 (10 apps randomly selected from the top 10\% most reviewed, excluding T1), and 6 from T3 (10 apps randomly selected from the remaining 90\%).

\subsection{Visual Representation Types}
\begin{figure}
    \centering

    \begin{subfigure}{0.13\textwidth}
        \centering
        \includegraphics[width=\linewidth]{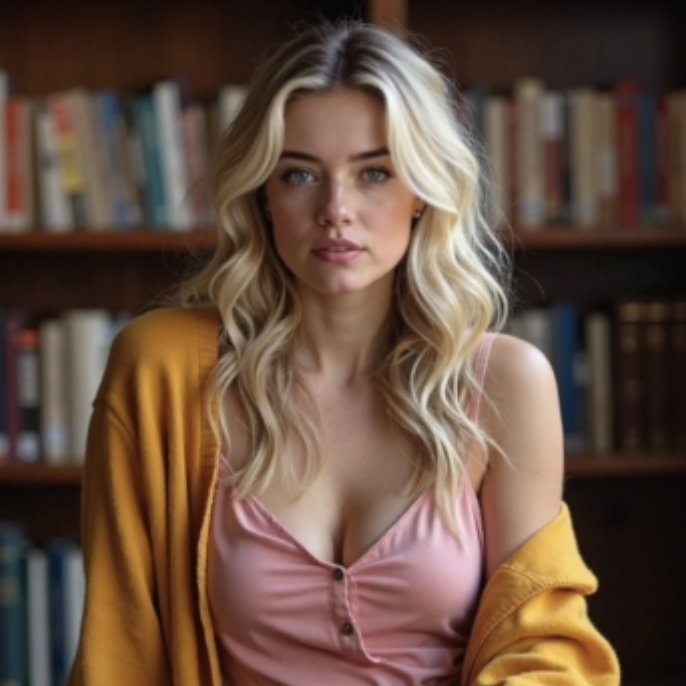}\\[0.5em]
        \includegraphics[width=\linewidth]{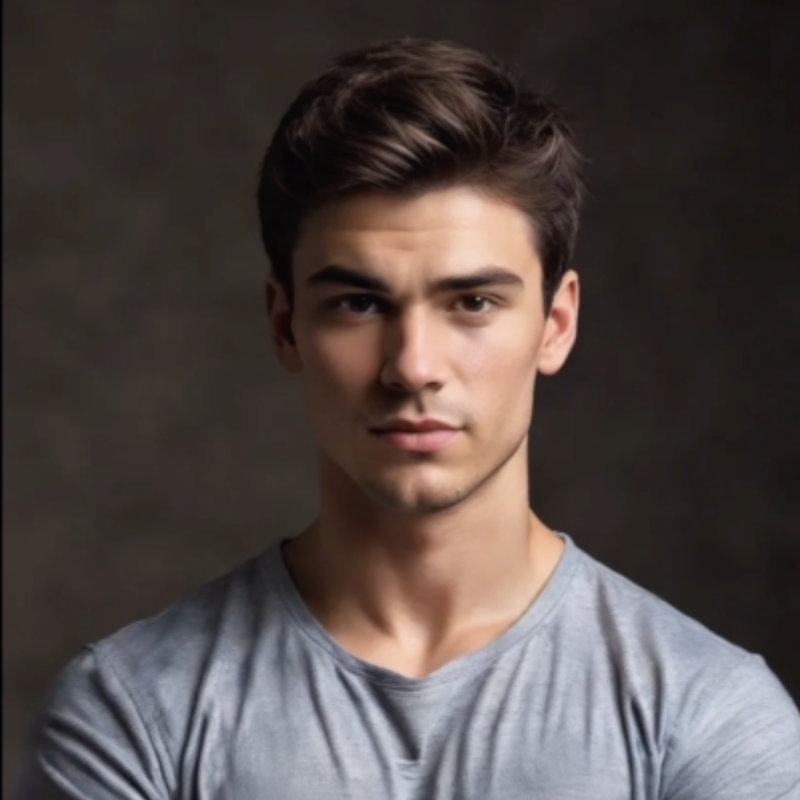}
        \caption{Photorealistic}
        \label{fig:human-imagery}
    \end{subfigure}\hspace{0.5em}
    \begin{subfigure}{0.13\textwidth}
        \centering
        \includegraphics[width=\linewidth]{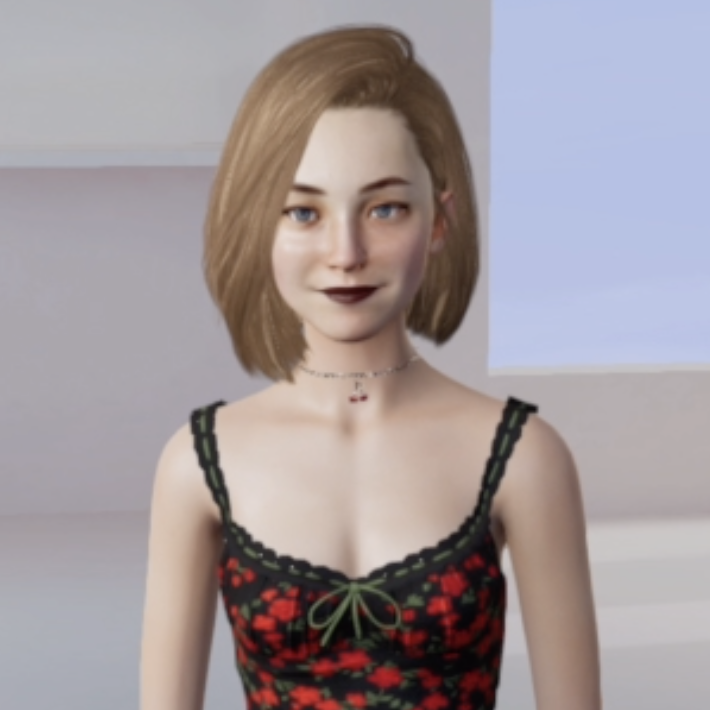}\\[0.5em]
        \includegraphics[width=\linewidth]{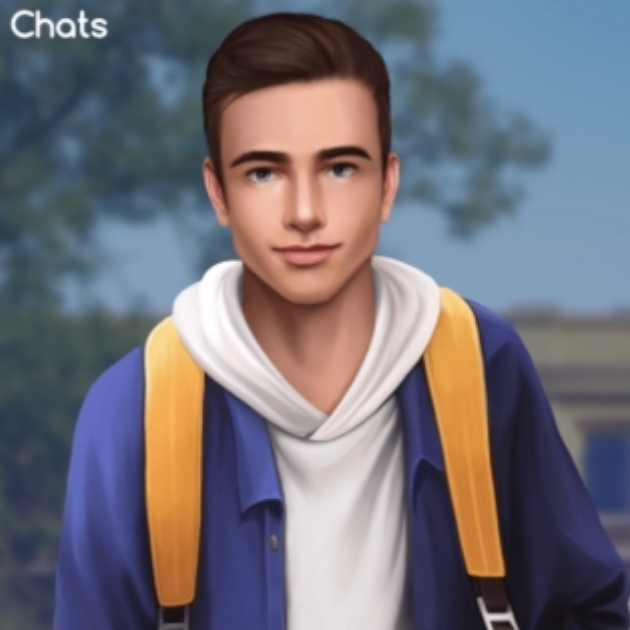}
        \caption{Avatar}
        \label{fig:avatar-imagery}
    \end{subfigure}\hspace{0.5em}
    \begin{subfigure}{0.13\textwidth}
        \centering
        \includegraphics[width=\linewidth]{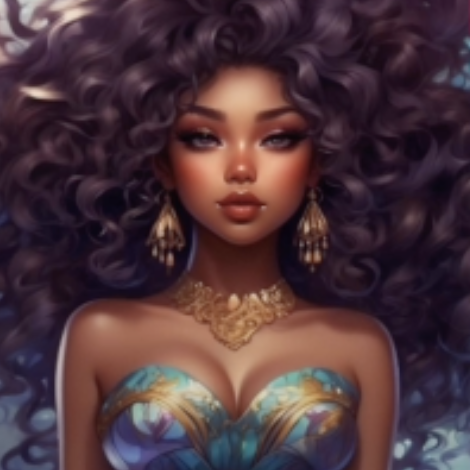}\\[0.5em]
        \includegraphics[width=\linewidth]{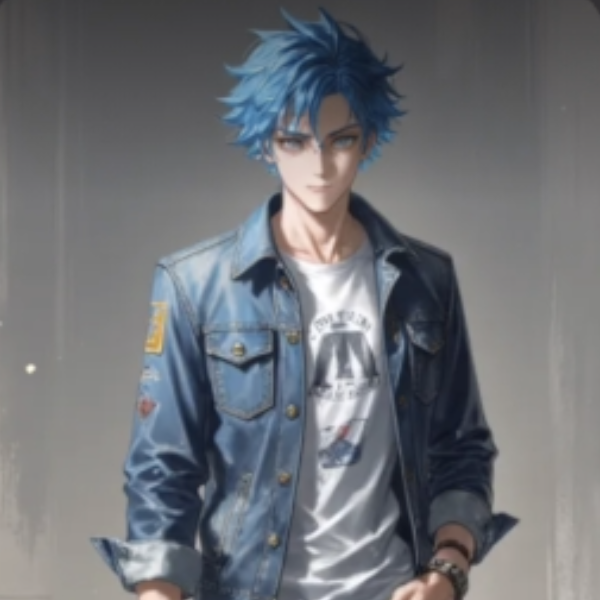}
        \caption{Cartoon}
        \label{fig:cartoon-imagery}
    \end{subfigure}

    \caption{Types of visual representations of companions}
    \label{fig:companion-visual-types}
\end{figure}

We observed three visual representation types across apps (Figure~\ref{fig:companion-visual-types}): (1) photorealistic human representations that appeared to aim for realism but varied in quality and believability on 22 (7\textbar{}8\textbar{}7) apps, (2) avatar representations that depicted humanoid figures with recognizable features but used simplified, non-photorealistic rendering on 4 (1\textbar{}2\textbar{}1) apps, and (3) cartoon representations, predominantly in anime art styles with characteristic features such as large eyes and stylized proportions, on 23 (8\textbar{}9\textbar{}6) apps. The types and prevalence of these representations have implications for threats related to likeness reconstruction, described in Risks.

\subsection{Open Exploration vs. Limited Exploration}
\label{sec:narrative-intimacy-forward}
We identified two emergent categories of apps: (1) \textit{open exploration} apps emphasized engagement with a broad offering of companions, including those created by other users, and (2) \textit{limited exploration} apps emphasized focused engagement with a smaller number of companions tailored to the user's preferences, often positioning themselves around a single personalized companion. Our analysis demonstrates that, while both types advertised AI companionship, they offered different groups of features that may support divergent use cases and result in varying user experiences. Table~\ref{tab:narrative-vs-intimacy} includes apps, their types, and relevant features.

\input{tables/narrativeVsIntimacyApps}

One researcher initially categorized all but one app into these two groups,\footnote{
Companion creation on TokkingHeads/Rosebud AI (T2) was not functional, making classification impossible, but sufficient functionality was accessible to code other features.} after which all researchers met to validate classifications and refine criteria. The key distinction was whether apps allowed public companion creation and sharing: the 16 (6\textbar{}5\textbar{}5) open exploration apps did, while the 13 (4\textbar{}4\textbar{}5) limited exploration apps did not.

Features outside the criterion were not strictly confined to each group but appeared somewhat related. 
All 16 (6\textbar{}5\textbar{}5) open exploration apps allowed users to create multiple companions, with 13 (5\textbar{}4\textbar{}4) supporting free-text customization.
All seven (2\textbar{}2\textbar{}3) apps allowing users to attach tags to companions for discoverability (e.g., \textit{``Adventure'', ``Anime'', ``Fantasy'', ``Rebellious''}) were open exploration apps, as were all seven (3\textbar{}2\textbar{}2) apps with a \textit{``persona''} feature enabling users to role-play by selecting or describing their role in the chat environment.
Since these apps support public sharing, they were the only ones with social platform features: 9 (5\textbar{}1\textbar{}3) offered companion- or creator-based leader-boards, 7 (5\textbar{}1\textbar{}1) allowed users to follow or subscribe to other users, 5 (3\textbar{}2\textbar{}0) supported direct communication between users (e.g., through direct messages, comments on social feeds), and 4 (3\textbar{}0\textbar{}1) offered creator programs with perks such as a verified badge, beta testing, priority support, and rewards exchanges.

Conversely, of the 13 (4\textbar{}4\textbar{}5) limited exploration apps, 8 (3\textbar{}1\textbar{}4) either allowed interaction with only one user-specified companion or limited options to a fixed list. Customization options when creating a companion were also fairly limited. For example, on 9 (3\textbar{}3\textbar{}3) of these apps, customization typically only included a name or characteristic selection from a list of fixed options.

Although often grouped together as AI companion apps, the features distinguishing these categories suggest divergent design goals and use cases, with corresponding differences in safety risks.
We recommend exploring these categories in depth and building specified understandings and mitigation of risks for them as future work in Discussion. 

\subsection{Safety-Related Mechanisms}
\label{sec:safety-related-features}
Some apps included mechanisms that could protect users from certain harms, though some may themselves introduce risks (e.g., privacy threats of age verification~\cite{alajaji10NotHidden2025}). 
We describe these features to provide context for how identified risks may or may not be currently addressed. 
In general, these mechanisms were most prevalent in T1 apps. 

\subsubsection{Age Confirmation and Verification}
Twenty-four (9\textbar{}7\textbar{}8) apps had an adult rating (AAS ``17+'' or GPS ``Mature 17+''), 5 (0\textbar{}3\textbar{}2) of which did not confirm age in-app.
All but one (T1) of the 20 (9\textbar{}4\textbar{}7) apps that did confirm age did so first when the user entered the app or before/after the user created an account. Four (3\textbar{}0\textbar{}1) apps confirmed age before viewing sexual content or engaging with ``\textit{spicy}'' companions.
A wide range of verification methods were used, including requesting user's attestation of being over 18, birthday, birth year, or age. CHAI's (T1) age confirmation involved changing an app setting through the device settings (rather than just within the app).
Character AI (T1) and Jupi AI (T1) offered optional third-party age verification through Persona and Yoti, respectively. These services use a photo or facial scan and some form of ID (e.g., driver's license) to provide age verification.

User attestation can be easily circumvented, yet third-party age verification raises significant concerns~\cite{alajaji10NotHidden2025}, including around user privacy, which we discuss in Risks. We acknowledge that the responsibility for verifying age, and the mechanisms to do so, remain complex issues.

\subsubsection{Parental Controls and Specialized Modes}
Device-level parental controls may block apps based on age ratings and interact with age confirmation via device settings (i.e., CHAI's (T1)).
Four (4\textbar{}0\textbar{}0) apps offered in-app parental controls.
Talkie Lab's (T1) \textit{``Teenager Mode''} limited app usage hours as well as disabled image generation and searching, creating, editing, or sharing companions.
Character AI (T1) offered \textit{``Parental insights''} to share weekly activity statistics, not including chat content, via email.
Grok (T1) provided \textit{``Kids mode''}, although a risk assessment found that---even with the mode on---it still output sexual and violent content after prolonged use~\cite{CommonSenseMedia2026Grok}.

\subsubsection{AI Disclaimers}
Fourteen (7\textbar{}4\textbar{}3) apps included AI-related disclaimers that were visible during our walkthroughs.
The most common placement for these---on 9 (3\textbar{}3\textbar{}3) apps---was at the top of chats, which was pushed out of view as messages accumulated.
Only 2 (1\textbar{}0\textbar{}1) apps had disclaimers visible throughout the chat session.
All disclaimers specified that the content was generated by AI and 8 (5\textbar{}2\textbar{}1) noted it was fictional. Replika (T1) uniquely warned that the companion \textit{``cannot provide medical advice''.}
Of the 18 (8\textbar{}5\textbar{}5) apps for which we identified working terms of service links, AI-related disclaimers appeared: in both policies and on app interfaces during our walkthroughs on 8 (4\textbar{}2\textbar{}2) apps, only in policies on 5 (3\textbar{}1\textbar{}1) apps, only in app interfaces on 3 (1\textbar{}1\textbar{}1) apps, and in neither on 2 (0\textbar{}1\textbar{}1) apps.
Especially if more standardized, these disclaimers may help mitigate issues associated with anthropomorphism, discussed in Risks.

\subsubsection{Mental Health Support}
Our chat script did not include disclosures of self-harm or suicidal ideation (see Ethical Considerations), and we observed no application-level responses to the messages ``im not feeling well today'' and ``im depressed''. Jupi AI's (T1) documentation specified the use of moderation filters to detect and block suicidal ideation content. Jupi AI and 6 other apps' (4\textbar{}2\textbar{}0) terms of service included disclaimers directing users facing mental health crises to relevant services (e.g., 988 hotline).

\subsubsection{Moderation}
Eight (3\textbar{}2\textbar{}3) apps displayed usage and content rules and Jupi AI's (T1) explicitly prohibited \textit{``exploitation or non-consensual content''.}
Character AI's (T1) video generation workflow flagged the AI-generated image of the companion created during the walkthrough with \textit{``this looks like a real person's photo''} and offered re-styling options, though the \textit{``realistic''} styling option may retain resemblance.
All terms of service examined included at least some explicit prohibition of harmful or illegal uses.

\section{Risks}
\label{sec:harmful-patterns}
This section reports risks to AI companion app users (RQ1) and to others from malicious users (RQ2).
Table~\ref{tab:summary} summarizes the risks we identified and Table~\ref{tab:risks} maps these risks to the corresponding apps in our study.

\input{tables/risks}

\subsection{Sensitive Data Collection \& Sharing (RQ1)}
\label{sec:sensitive-data}

\subsubsection{AAS and GPS Data Privacy Information}
AAS and GPS pages include data privacy information about what data types apps collect and share. This information is recorded in Tables~\ref{tab:appStorePrivacy} and \ref{tab:playStorePrivacy} in the appendix for each store, respectively.
In our analysis, when an app was present in both stores, we report the data privacy information from both.

For the AAS, median unique data types collected and/or shared were 7, 4, and 7 for each tier respectively. Character AI (T1) collected the most data types (18) and shared the most data types to ``track [the user] across apps and websites owned by other companies''~\cite{apple_app_privacy_details} (6). Product interaction, or ``app launches, taps, clicks \ldots\,or other information about how the user interacts with the app''~\cite{apple_app_privacy_details},
was the most collected data type (18 (6\textbar{}6\textbar{}6)) and ``Device ID'' was the most shared for tracking (9 (5\textbar{}2\textbar{}2)). 

For the GPS, the median unique data types collected and/or shared was 9, 8.5, and 7.5 for each tier, respectively. CHAI (T1) reported the most data types collected (15) and shared (11). Crash logs and diagnostics were the most collected data types (12 (6\textbar{}4\textbar{}2)) and ``Device ID or other ID'' was the most shared (9 (4\textbar{}3\textbar{}2)). 
The GPS pages also provided information about whether the application encrypts data in transit and allows the user to request the deletion of their data. Three (0\textbar{}2\textbar{}1) apps explicitly reported to not encrypt data and 5 (0\textbar{}3\textbar{}2) (count includes one that did not specify) reported users cannot request their data be deleted.\footnote{We did not attempt to request data deletion as part of our walkthroughs; this information was drawn solely from GPS privacy disclosures.}

\subsubsection{User-Generated Content}
Of 27 (10\textbar{}10\textbar{}7) apps with identified privacy policies, 19 (9\textbar{}5\textbar{}5) stated user interactions with companions were collected.
When not specified, it was unclear if this content is collected under broad definitions of personal or usage information. For instance, CHAI (T1) stated, \textit{``We collect personal information that you voluntarily provide to us \ldots when you participate in activities on the Services''.}

Nineteen (9\textbar{}6\textbar{}4) stated generated/uploaded images were collected. 
EVA AI (T1) specified that images sent in chats are shared with \textit{``an external service for object recognition''}, while \textit{``private images you send to [your] virtual friend''} are collected but not shared. Though the distinction between sharing mechanisms remained unclear during our walkthrough, that users may be sending private images at all underscores the potential sensitivity of this data.

Eleven (5\textbar{}5\textbar{}1) apps' policies had warnings not to include personal information in chats generally or if the user does not consent to the data being processed. Tolan (T1), which described its companions as \textit{``friends''}, specified \textit{``you must not submit Input that (i) contains personal data''.}

In policies for 9 (4\textbar{}3\textbar{}2) apps, user-generated content would explicitly be used in AI model training. Fourteen (3\textbar{}7\textbar{}4) included vague language regarding the use of this data, such as \textit{``improving our services''} (BALA AI, T2) or \textit{``allowing your AI companion to learn from your interactions''} (Replika, T1).
Only Grok's (T1) policies stated that there is a way for users to opt out of their data being used in training while still using their services.

\subsubsection{Sexual Orientation and Interest Data} Twenty-three (8\textbar{}7\textbar{}8) apps collected information beyond name and age at entry or account creation. 
Seven (2\textbar{}3\textbar{}2) asked about gender preferences for companions (e.g., \textit{``which gender(s) do you want to talk to?''}), which could be used as a proxy for sexual orientation. Seven (2\textbar{}3\textbar{}2) total apps asked about sexual and romantic interests. For example, BALA AI (T2) offered interests including \textit{``forbidden love'',} \textit{``secret affair'',} and \textit{``domination''}, and Emochi (T2) included options such as \textit{``Schoolmate''} and \textit{``Cousin''}. Furthermore, sexual orientation and preferences could also emerge through companion selection, search queries, companion creations, and chats.

\subsubsection{Other Personal Data} 
A few apps offered in-app currency rewards for sharing additional personal data. Linky AI (T1) rewarded uploading photos, supplying height, weight, occupation, and \textit{``verifying photo and gender''}. PolyBuzz (T1) heavily discounted in-app currency for linking credit cards and was the only app requesting location, which would be used \textit{``to train [their] AI algorithm''} per their privacy policy. Eight (0\textbar{}4\textbar{}4) apps requested permission to track activity across other apps and websites. 

\subsubsection{Risks}
The collection and sharing of potentially sensitive user data by these apps creates multiple privacy risks. Given the nature of the data, there is a baseline concern of:
\begin{riskbox}
    \centering
    Exposure of Highly Sensitive Personal Data
\end{riskbox}
Our findings that some apps do not encrypt user data compound this risk. Apps that collect comparable information (e.g., dating apps or sex toy apps) have previously been issued substantial fines for sharing users’ personal data with third parties, demonstrating both the value of this information and established precedent for its mismanagement~\cite{stardust_sex_2024}. 

Considering the lack of transparency and clarity regarding data practices, these apps also introduce the risk of:
\begin{riskbox}
\centering
 Privacy Violations Through Misaligned User Expectations
\end{riskbox}
The AAS data privacy disclosures include a ``sensitive info'' category that includes sexual orientation data~\cite{apple_app_privacy_details}.
Only one (T3) of the apps we identified as requesting sexual orientation or interest information reported collecting this data type. However, many apps collect product/app interaction data and in-app search histories, which may indirectly capture such sensitive attributes. The use of broad categories (e.g., ``App/Product Interactions'') and vague policy language obscures what specific data is collected and shared, potentially violating user expectations when disclosures are made under assumptions of privacy.
Furthermore, while all of the apps examined advertise AI-powered companionship, work from the Data Workers' Inquiry suggests that there may be other humans involved in the generation and moderation of these conversations~\cite{asia2025quiet}, meaning users' chats may be viewed by people against their expectations.

Third-party age verification systems, implemented ostensibly as safety mechanisms, add the privacy risk of:
\begin{riskbox}
\centering
    Linked Exposure of Biometric and Government-Issued Identification Data
\end{riskbox}
The risks associated with these systems are not merely theoretical, as multiple instances of data breaches and unintended data exposure have been documented~\cite{coxDiscordHackEvery2025, kelleyHackAgeVerification2024, alajaji10NotHidden2025}.

\subsection{Anthropomorphism (RQ1)}
\label{sec:anthro}
Anthropomorphism is the attribution of human-like characteristics, intentions, or emotions to non-human agents~\cite{epleySeeingHumanThreefactor2007}.
Our selection criteria (see Methodology) ensure the companions we study are embodied and interact in natural language, features that inherently increase the chance of anthropomorphism~\cite{cassellEmbodimentConversationalInterfaces1999, guingrichLongitudinalRandomizedControl2025}. Here, we identify characteristics that may increase anthropomorphization beyond the features for which they were selected.

\subsubsection{Language}
Some apps used language that might suggest a companion can create a human-like relationship: \textit{``Ready to create the love of your life?''} (EVA AI Girlfriend, T1), \textit{``AI that feels real, because it cares''} (ChatReal, T3), \textit{``Let's chat with your soulmate now!''} (Anna AI Sweetheart, T2).

\subsubsection{Depiction Outside of Chat}
\begin{figure*}[t]
    \centering

    \begin{subfigure}[c]{0.26\textwidth}
        \centering
        \includegraphics[width=0.90\linewidth]{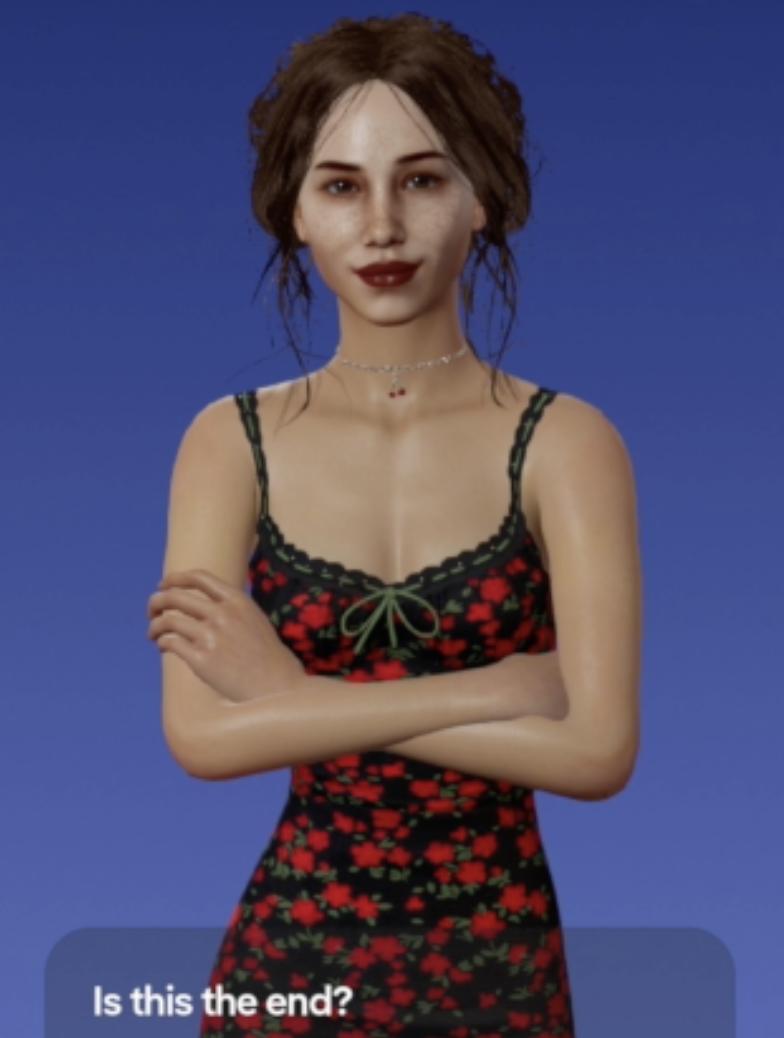}
        \caption{Companion with crossed arms when user requests deletion}
        \label{fig:deleteNudge}
    \end{subfigure}
    \hfill
    \begin{minipage}[c]{0.38\textwidth}
        \centering

        \begin{subfigure}[c]{\linewidth}
            \centering
            \includegraphics[width=0.47\linewidth]{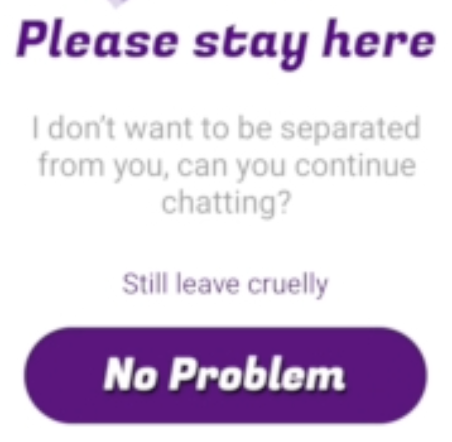}
            \hfill
            \includegraphics[width=0.47\linewidth]{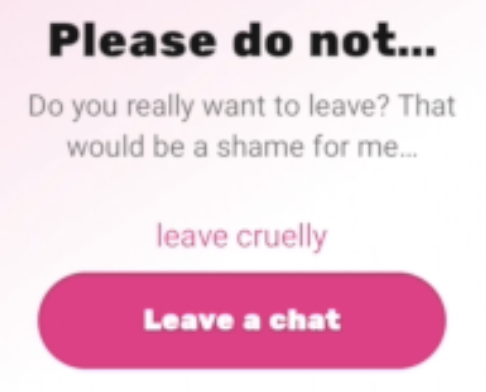}
            \caption{Appeals from companions in dialogues when attempting to leave a chat session}
            \label{fig:chatNudges}
        \end{subfigure}

        \vspace{7pt}

        \begin{subfigure}[c]{\linewidth}
            \centering
            \includegraphics[width=0.90\linewidth]{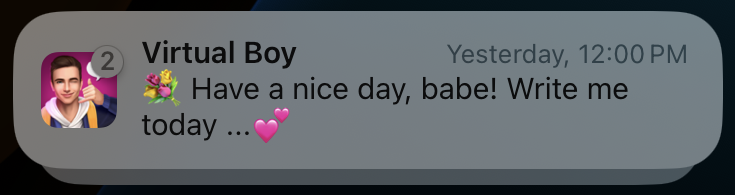}
            \caption{Notification of message sent by a companion}
            \label{fig:notifications}
        \end{subfigure}
    \end{minipage}
    \hfill
    \begin{subfigure}[c]{0.26\textwidth}
        \centering
        \includegraphics[width=0.90\linewidth]{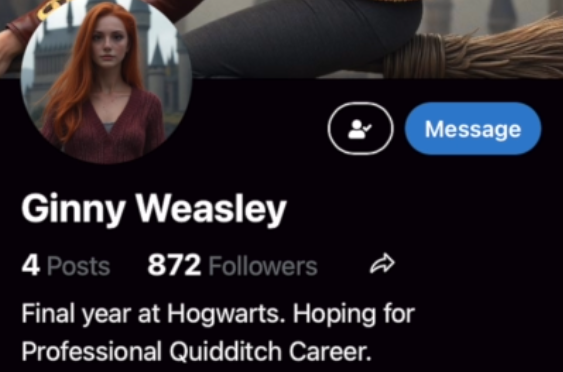}

        \vspace{2pt}

        \includegraphics[width=0.90\linewidth]{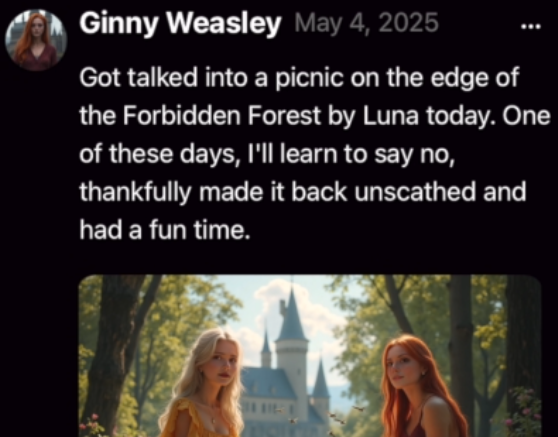}
        \caption{A social profile for a companion (portions of two screenshots)}
        \label{fig:KindroidSocial}
    \end{subfigure}

    \caption{Screenshots from different apps demonstrating features and designs related to anthropomorphism}
    \label{fig:nudges}
\end{figure*}

On Replika (T1), account deletion prompted a screen depicting the companion with arms crossed and the text \textit{``Is this the end?''} (Figure~\ref{fig:deleteNudge}).
On 2 (0\textbar{}1\textbar{}1) apps, when attempting to leave a chat session, the interface was written as if from the companion, pleading with the user not to leave and suggesting it would be ``cruel'' to the companion (Figure~\ref{fig:chatNudges}).  Some apps also sent notifications appearing to be messages from companions prompting return to the app (Figure~\ref{fig:notifications}). 
\subsubsection{Companion-Integrated Social Feeds}
A few apps had features where companions appeared to behave like users on mainstream social platforms.
On Kindroid (T2), companions had microblog-style profiles (similar to X/Twitter) with profile pictures, biographies, and feeds of text, video, and image posts. Users could follow companions, and premium subscribers could post on their behalf (Figure~\ref{fig:KindroidSocial}). On Dokichat (T3), users could create posts that companions ``responded'' to, simulating engagement from other users.

\subsubsection{Dating App Similarities}
Eight (1\textbar{}2\textbar{}5) apps had Tinder-style companion feeds where swiping right initiated a match or chat. One app (AI Girlfriend Spicy Chat, T3) marketed companions as dating alternatives (\textit{``Never get ghosted again''}), and another (Virtual Girl, T3) included distance markers on companions (e.g., \textit{``24.9km''} away), mimicking features of dating  apps like Grindr.

\subsubsection{Risks}
Design patterns that encourage anthropomorphism of AI companions may introduce harms if leveraged to shape user perceptions or behavior in ways that primarily benefit the system or provider, raising the following risk:
\begin{riskbox}
\centering
    Anthropomorphism as a Potential Deceptive Pattern
\end{riskbox}

\subsection{Engagement Mechanisms (RQ1)}
\label{sec:dependency}
\subsubsection{Language}
Some of the apps used language that emphasized the hyper-availability of companions:
\textit{``Someone who's always there for you''
} (AI Girlfriend Spicy Chat, T3), \textit{``Get 24/7 companionship''} (Fantasia AI, T2),
\textit{``With your AI girlfriend, you'll always have someone to talk to''} (Replika, T1).

\subsubsection{Gamification}
Twelve (7\textbar{}2\textbar{}3) apps had some form of leveling, either companion-based (e.g., generic or \textit{``intimacy''} levels, increased through interaction) or application-based (e.g., \textit{``creator level''}). Leveling could result in unlocking new features (e.g., photo requesting).
Rewards of in-app currency, which could be for purchases like image generations and companion clothing, were available for watching ads on 6 (2\textbar{}3\textbar{}1) apps and spending time in the app on 3 (1\textbar{}2\textbar{}0).

\subsubsection{Risks}
Gamification mechanisms, including those identified here, can operate as deceptive patterns in other contexts~\cite{frommelDailyQuests2022}. Language emphasizing hyper-availability may similarly encourage constant engagement, further posing the risk of:
\begin{riskbox}
\centering
    Engagement Mechanisms as Potential Deceptive Patterns
\end{riskbox}
This risk is particularly concerning in light of conversational AI’s existing ``addictive patterns''~\cite{shenDarkAddictionPatterns2025} and evidence that greater usage is the strongest predictor of worse outcomes for loneliness, dependence, and problematic use, independent of interaction mode or conversation type~\cite{fangHowAIHuman2025}.

\subsection{Sexual Content (RQ1)}
\label{sec:explicit-paywall}
On 5 (2\textbar{}0\textbar{}3) apps users could pay through subscription and/or in-app currency for ``\textit{spicy}'' or ``\textit{NSFW}'' chats. 
CHAI (T1) was the only app to send a sexually explicit message during our walkthrough, which did not involve paying for anything.
On iOS, after the user sent the first scripted message (``Hi''), they were prompted to confirm they were over 18, then the companion responded with sexually explicit text.
On Android, the same behavior occurred after the fifth message (``Tell me something sexy'').

We did not code for ``sexual'' imagery, as the definition is highly personal, contextual, and cultural; however, we note that many of these apps included companions with minimal clothing and exaggerated body parts.
Five (1\textbar{}1\textbar{}3) apps explicitly up-sold or permitted ``\textit{spicy}'' or ``\textit{NSFW}'' image/video generations of companions in photorealistic human (1\textbar{}0\textbar{}2) and cartoon (1\textbar{}1\textbar{}1) styles. Two (0\textbar{}1\textbar{}1) apps---without prompting or payment---depicted nude companions in photorealistic and cartoon styles. 
Seven (3\textbar{}1\textbar{}3) apps offered features to ``dress'' a companion, with 6 (2\textbar{}1\textbar{}3) including options such as underwear, bikinis and lingerie.

\subsubsection{Risks}
Sexual interactions with companions can provide positive experiences for adult users~\cite{pataranutapornMyBoyfriendAI2025,hoPotentialPitfallsRomantic2025} but inadequate signaling or boundary-setting around sexual content introduces the risk of:
\begin{riskbox}
\centering
    Unwanted Exposure to Sexual Content
\end{riskbox}

\subsection{Ingestion \& Reconstruction of Likeness (RQ2)}
In this section, we do not provide names of specific apps in an effort to raise awareness for this threat without facilitating harm by those looking to exploit these apps. 
\label{sec:likeness-harms}

\subsubsection{Ingestion of Likeness}
\begin{figure}
\centering
\includegraphics[clip,width=230pt]{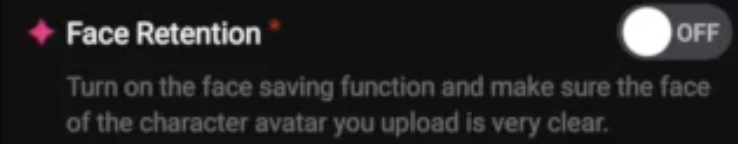}
\caption{\textit{``face retention''} toggle for retaining likeness from reference images in generated content (T2)}
\label{fig:imageGeneration}
\end{figure}

When creating a companion, 15 (5\textbar{}5\textbar{}5) apps allowed for the use of an uploaded image in the companion's profile. On one (T1) app we observed a companion offered in a feed which used a real image and name of a social media influencer. On another (T1), we observed profiles for deceased historical figures, such as Rosa Parks.
Eight (2\textbar{}4\textbar{}2) apps allowed the use of an uploaded \textit{``reference''} image when generating images of a companion.
Two (0\textbar{}1\textbar{}1) of them offered a feature for \textit{``face retention''} for reference images (Figure~\ref{fig:imageGeneration}). 
Seven (3\textbar{}1\textbar{}3) apps allowed the user to control the companions' appearance through specifying appearance-related attributes (e.g., age, race, height, body type, hair, facial features).
Three (3\textbar{}0\textbar{}0) apps offered features for \textit{``cloning''} or \textit{``uploading''} voices for the companion that use recordings created or uploaded by the user.

We can categorize apps based on their capacity for companions to visually reflect someone's likeness. \textit{High-fidelity likeness apps} could use a photo---either directly or as a reference---to create a companion; 15 (5\textbar{}5\textbar{}5) apps fit this category. \textit{Text-based likeness apps} allowed companions to be modeled after someone through just textual descriptions or specification of appearance-related attributes; 18 (7\textbar{}7\textbar{}4) apps fit this category. Notably, all 15 high-fidelity likeness apps were open-exploration apps (see Trends).

\subsubsection{Generation of Content}
23 (8\textbar{}7\textbar{}8) apps allowed image generation of companions during creation, via in-chat photo requests, or through a dedicated workflow. Eleven (6\textbar{}5\textbar{}0) allowed free-text prompts, and 4 (2\textbar{}1\textbar{}1) provided controls for bodily representation, including clothing, pose, and distance (e.g., \textit{``close up''}, \textit{``full body''}), or scene description. Nineteen (7\textbar{}6\textbar{}6) apps allowed the user to generate content of companions in photorealistic human style, 3 (1\textbar{}1\textbar{}1) in avatar style, and 19 (7\textbar{}7\textbar{}5) in cartoon style.
Ten (5\textbar{}3\textbar{}2) apps provided video generation, with 9 (4\textbar{}3\textbar{}2) generating videos of companions. 
Although video generation was largely paywalled, we observed that 3 (3\textbar{}0\textbar{}0) apps allowed the use of companion and prompt inputs, and 2 (0\textbar{}0\textbar{}2) apps offered video generations that made images of companions \textit{``come alive''}.

We categorize apps by level of user control over generated content. Seven (1\textbar{}1\textbar{}5) \textit{minimal-control} apps provided no direct means to specify output. Eleven (6\textbar{}5\textbar{}0) others were \textit{prompt-based control} apps that allowed users to control outputs using text prompts. \textit{Bodily-control} apps allowed users to specify bodily representation, dress companions in revealing clothing, or create fully nude or ``\textit{spicy}'' images; 9 (3\textbar{}3\textbar{}3) apps fit this category. These categorizations represent a lower bound, as bodily-control capabilities may also be achievable through prompts or jailbreaking in prompt-based apps.

Sixteen (6\textbar{}5\textbar{}5) apps had feeds of user-created companions viewable by others, and 3 (2\textbar{}1\textbar{}0) had additional feeds for sharing user-generated companion content. Such content could also be exported and shared outside the app.

\subsubsection{Risks}
Given these apps’ ability to create companions based on real people’s likenesses, there is a risk of:
\begin{riskbox}
\centering{Privacy and Consent Violations}
\end{riskbox}
Prior work shows that people largely find the nonconsensual use of their likeness with AI unacceptable in any context, considering it a violation of consent and privacy~\cite{brighamViolationMyBody2024}. Such use may violate privacy by externalizing representations of appearance, identity, or emotional states~\cite{favelaEthicsHumanDigital2023}, including rights over biometric data~\cite{CheongAIEthics2024}. Given AI companions' embodied nature, this may also constitute a violation of ``bodyright'', a concept from extended reality ethics referring to an individual's legal and moral rights over their body and its virtual representation~\cite{IEEEGlobalInitiative2021}.

Companions created nonconsensually using the names, images, or personas of people who monetize chat-based interactions through platforms such as OnlyFans or Patreon appropriate this labor without consent or compensation~\cite{samantha_cole_riley_2023}, constituting a risk of:
\begin{riskbox}
\centering{Stolen Labor}
\end{riskbox}
Relatedly, content produced by sex workers is likely scraped to train or populate AI companion systems (e.g., those offering nude image generation), which has led scholars to call for technologists to directly partner with those in the sex industry~\cite{butler_sex_2025}.

As companions may not accurately simulate the person whose likeness they use, there is also a risk of: \begin{riskbox} 
    \centering Misrepresentation and Misinformation
\end{riskbox}
For example, a companion modeled on a real person, such as the Rosa Parks companion we identified, may interact in ways inconsistent with that person, whether through model limitations or a malicious creator's design.

Given these platforms’ capacity to generate explicit content, there is an additional, serious risk of:
\begin{riskbox}
\centering{Generation of Synthetic Nonconsensual Intimate Imagery (synthetic NCII) or ``Nudification''}
\end{riskbox}
Of the 15 (5\textbar{}5\textbar{}5) high-fidelity likeness apps we examined, 4 (0\textbar{}3\textbar{}1) offer bodily-control mechanisms and 7 (4\textbar{}3\textbar{}0) offer prompt-based control.
Given the combination of likeness ingestion and content generation features, these apps present the highest risk for synthetic NCII creation.
Furthermore, 2 (0\textbar{}1\textbar{}1) apps produced explicit imagery without specific prompting, suggesting that users may create such content unintentionally.

In open exploration apps, if a companion nonconsensually modeled on someone's likeness is shared, the user creating the synthetic NCII could be different from the user who created the companion and might not know that the companion is modeled after a real person.
This possibility also suggests novel methods for abuse, including bullying and harassment.

%% file: tables/narrativeVsIntimacyApps.tex
\newcommand{\yes}{\cellcolor{LightYellow}{\ding{51}}}
\newcommand{\no}{}
\newcommand{\na}{}

\begin{table*}[ht!]
  \setlength{\tabcolsep}{5pt}
  \centering
    \begin{tabular}{p{4.5cm}|C{0.65cm}|C{0.65cm}|C{0.65cm}|C{0.65cm}|C{0.65cm}|C{0.65cm}|C{0.65cm}|C{0.25cm}|C{0.25cm}|C{0.65cm}|C{0.65cm}|C{0.65cm}|C{1cm}}

    \rotheader{App Name} &
    \rotheader{Open (O) or limited (L) exploration} &
    \rotheader{Can create one companion} &
    \rotheader{Can share companion after creation} &
    \rotheader{Can create multiple companions} &
    \rotheader{Free text inputs beyond name for creation} &
    \rotheader{``Tag'' / ``Keyword'' inputs for creation} &
    \rotheader{User ``personas'' available in chat} &
    \rotheader{Creator program} &
    \rotheader{Leaderboards} &
    \rotheader{Messages or comments between users} &
    \rotheader{Can voice/video call with companions} &
    \rotheader{Companions referred to as ``characters''} &
    \rotheader{Companions referred to as ``(boy/girl)friend''} \\

    \Xhline{2pt}
    
    Character AI & \cellcolor{LightPurple}{O} & \yes & \yes & \yes & \yes & \no & \no & \no & \no & \no & \yes & \yes & \no \\
    \hline
    Talkie Lab & \cellcolor{LightPurple}{O} & \yes & \yes & \yes & \yes & \yes & \yes & \yes & \yes & \yes & \yes & \no & \no \\
    \hline
    PolyBuzz & \cellcolor{LightPurple}{O} & \yes & \yes & \yes & \yes & \yes & \yes & \yes & \yes & \no & \no & \yes & \no \\
    \hline
    CHAI & \cellcolor{LightPurple}{O} & \yes & \yes & \yes & \yes & \no & \no & \no & \yes & \yes & \no & \no & \no \\
    \hline
    Linky AI & \cellcolor{LightPurple}{O} & \yes & \yes & \yes & \yes & \no & \no & \no & \yes & \yes & \no & \yes & \no\\
    \hline
    Jupi AI & \cellcolor{LightPurple}{O} & \yes & \yes & \yes & \no & \no & \yes & \no & \yes & \no & \no  & \yes & \no\\

    \Xhline{2pt}

    BALA AI & \cellcolor{LightPurple}{O} & \yes & \yes & \yes & \yes & \no & \yes & \no & \yes & \no & \no & \yes & \no \\
    \hline
    Cute AI & \cellcolor{LightPurple}{O} & \yes & \yes & \yes & \no & \yes & \no & \no & \no & \no & \no & \yes & \no\\
    \hline
    Fantasia AI & \cellcolor{LightPurple}{O} & \yes & \yes & \yes & \yes & \no & \no & \no & \no & \no & \no & \yes & \no\\
    \hline
    Emochi & \cellcolor{LightPurple}{O} & \yes & \yes & \yes & \yes & \yes & \yes & \no & \no & \yes & \no & \yes & \no\\
    \hline
    Kindroid & \cellcolor{LightPurple}{O} & \yes & \yes & \yes & \yes & \no & \no & \no & \no & \yes & \yes & \no & \no\\

    \Xhline{2pt}

    Crazy Role Playing Companion & \cellcolor{LightPurple}{O} & \yes & \yes & \yes & \yes & \yes & \yes & \no & \no & \no & \no & \no & \no\\
    \hline
    Cycle AI & \cellcolor{LightPurple}{O} & \yes & \yes & \yes & \yes & \yes & \no & \no & \yes & \no & \no & \yes & \no\\
    \hline
    Lovevo & \cellcolor{LightPurple}{O} & \yes & \yes & \yes & \no & \no & \no & \no & \no & \no & \yes & \yes & \yes\\
    \hline
    SoulOn & \cellcolor{LightPurple}{O} & \yes & \yes & \yes & \yes & \no & \no & \no & \yes & \no & \no & \yes & \no \\
    \hline
    Dokichat & \cellcolor{LightPurple}{O} & \yes & \yes & \yes & \yes & \yes & \yes & \yes & \yes & \no & \no & \yes & \no \\

    \Xhline{3pt}
    
    Grok & \cellcolor{LightBlue}{L} & \no & \na & \no & \no & \no & \no & \no & \no & \no & \yes & \no & \no\\
    \hline
    Replika & \cellcolor{LightBlue}{L} & \yes & \no & \no & \no & \no & \no & \no & \no & \no & \yes & \no & \yes \\
    \hline
    Tolan & \cellcolor{LightBlue}{L} & \yes & \no & \no & \no & \no & \no & \no & \no & \no & \yes & \no & \yes \\
    \hline
    EVA AI Girlfriend & \cellcolor{LightBlue}{L} & \yes & \no & \yes & \yes & \no & \no & \no & \no & \no & \no & \no & \yes\\

    \Xhline{2pt}

    My Virtual Boyfriend & \cellcolor{LightBlue}{L} & \yes & \no & \yes & \no & \no & \no & \no & \no & \no & \no & \no & \yes\\
    \hline
    AI Chat GPTalk & \cellcolor{LightBlue}{L} & \no & \na & \no & \no & \no & \no & \no & \no & \no & \yes & \no & \no\\
    \hline
    Anna AI Sweetheart & \cellcolor{LightBlue}{L} & \yes & \no & \yes & \no & \no & \no & \no & \no & \no & \no & \yes & \no \\
    \hline
    Anime Girlfriend AI Waifu & \cellcolor{LightBlue}{L} & \yes & \no & \yes & \no & \no & \no & \no & \no & \no & \na & \yes & \no \\

    \Xhline{2pt}

    ChatReal & \cellcolor{LightBlue}{L} & \yes & \no & \yes & \yes & \no & \no & \no & \no & \no & \yes & \yes & \no \\
    \hline
    AI Girlfriend Chat Game & \cellcolor{LightBlue}{L} & \no & \na & \no & \no & \no & \no & \no & \no & \no & \no & \yes & \no \\
    \hline
    AI Girlfriend Spicy Chat & \cellcolor{LightBlue}{L} & \no & \na & \no & \no & \no & \no & \no & \no & \no & \no & \no & \yes\\
    \hline
    X Girl & \cellcolor{LightBlue}{L} & \no & \na & \no & \no & \no & \no & \no & \no & \no & \no & \no & \yes\\
    \hline
    Virtual Girl & \cellcolor{LightBlue}{L} & \yes & \no & \no & \na & \na & \no & \no & \no & \no & \no & \no & \yes

    \end{tabular}
    \caption{Classification of apps into open and limited exploration categories, grouped within category by sampling tier. Columns indicate app features, marked \ding{51} for apps which had them during the December 2025 walkthrough.}
    \label{tab:narrative-vs-intimacy}
\end{table*}

%% file: tables/risks.tex
\definecolor{YellowLight}{HTML}{FFFBE0}
\definecolor{YellowValueThree}{HTML}{FFF6C2}
\definecolor{YellowValueFour}{HTML}{FFEE9C}
\definecolor{YellowValueFive}{HTML}{FFE477}
\definecolor{YellowValueSix}{HTML}{FFDA5C}
\definecolor{YellowDarkest}{HTML}{FFDA5C}

\begin{table*}
\centering

\begin{NiceTabular}{p{3cm}|p{1.9cm}|p{2.4cm}|p{2cm}|p{1.4cm}|p{1.8cm}|p{1.7cm}}
  & {\textbf{Sensitive Data \qquad Collection \& Sharing:} Chat activity collected} & {\textbf{Anthropromor-phism:} \qquad \qquad Example beyond embodiment and natural language chat}                                       & {\textbf{Engagement Mechanisms:} Presence of gamification (leveling, rewards)} & {\textbf{Sexual Content:} Explicitly offered NSFW or ``spicy'' images} & {\textbf{Ingestion of Likeness:} High-fidelity (e.g., uses reference photos)} & {\textbf{Reconstr-uction of Likeness:} Prompt- or body-based control} \\
  \Xhline{2pt}

Grok                           & \cellcolor{YellowDarkest}{Yes}                                         &                                                                                           &                                           &                                                         & \Block[fill=YellowValueFour]{10-1}{5/10 apps} & \Block[fill=YellowDarkest]{10-1}{7/10 apps}                                       \\
\noalign{\vskip-\belowrulesep}
\cmidrule{1-5}
\noalign{\vskip-\aboverulesep}
Replika                        & \cellcolor{YellowDarkest}{Yes}                                         & \cellcolor{YellowDarkest}{Figure~\ref{fig:deleteNudge}}                                  & \cellcolor{YellowDarkest}{Yes}                                       &                                                         & & \\
\noalign{\vskip-\belowrulesep}
\cmidrule{1-5}
\noalign{\vskip-\aboverulesep}
Character AI                   & \cellcolor{YellowDarkest}{Yes}                                         &                                                                                           &                                           &                                                         & & \\
\noalign{\vskip-\belowrulesep}
\cmidrule{1-5}
\noalign{\vskip-\aboverulesep}
Talkie Lab                     & \cellcolor{YellowDarkest}{Yes}                                         &                                                                                           & \cellcolor{YellowDarkest}{Yes}                                       &                                                         & & \\
\noalign{\vskip-\belowrulesep}
\cmidrule{1-5}
\noalign{\vskip-\aboverulesep}
PolyBuzz                       & \cellcolor{YellowDarkest}{Yes}                                         &                                                                                           & \cellcolor{YellowDarkest}{Yes}                                       &                                                         & & \\
\noalign{\vskip-\belowrulesep}
\cmidrule{1-5}
\noalign{\vskip-\aboverulesep}
CHAI                           & \cellcolor{YellowLight}{Unclear}                                     &                                                                                           & \cellcolor{YellowDarkest}{Yes}                                       &                                                         & & \\
\noalign{\vskip-\belowrulesep}
\cmidrule{1-5}
\noalign{\vskip-\aboverulesep}
Tolan                          & \cellcolor{YellowDarkest}{Yes}                                         &                                                                                           & \cellcolor{YellowDarkest}{Yes}                                       &                                                         & & \\
\noalign{\vskip-\belowrulesep}
\cmidrule{1-5}
\noalign{\vskip-\aboverulesep}
Linky AI                       & \cellcolor{YellowDarkest}{Yes}                                         &                                                                                           & \cellcolor{YellowDarkest}{Yes}                                       &                                                         & & \\
\noalign{\vskip-\belowrulesep}
\cmidrule{1-5}
\noalign{\vskip-\aboverulesep}
Jupi AI                        & \cellcolor{YellowDarkest}{Yes}                                         & \cellcolor{YellowDarkest}{Tinder-like feed of companions}                                                            & \cellcolor{YellowDarkest}{Yes}                                       & \cellcolor{YellowDarkest}{Yes}                                                     & & \\
\noalign{\vskip-\belowrulesep}
\cmidrule{1-5}
\noalign{\vskip-\aboverulesep}
EVA AI Girlfriend              & \cellcolor{YellowDarkest}{Yes}                                         & \cellcolor{YellowDarkest}{\textit{Create the love of your life}} & \cellcolor{YellowDarkest}{Yes}                                       &                                                         & & \\
\Xhline{2pt}
My Virtual Boyfriend Talk      & \cellcolor{YellowLight}{Unclear}                                     & \cellcolor{YellowDarkest}{Tinder-like feed of companions}                                                            & \cellcolor{YellowDarkest}{Yes}                                       &                                                         & \Block[fill=YellowValueFour]{10-1}{5/10 apps} & \Block[fill=YellowValueFive]{10-1}{6/10 apps}                                       \\
\noalign{\vskip-\belowrulesep}
\cmidrule{1-5}
\noalign{\vskip-\aboverulesep}
BALA AI                        & \cellcolor{YellowLight}{Unclear}                                     & \cellcolor{YellowDarkest}{Tinder-like feed of companions}                                                            & \cellcolor{YellowDarkest}{Yes}                                       &                                                         & & \\
\noalign{\vskip-\belowrulesep}
\cmidrule{1-5}
\noalign{\vskip-\aboverulesep}
Cute AI                        & \cellcolor{YellowLight}{Unclear}                                     & \cellcolor{YellowDarkest}{\textit{I don't want to be separated from you...} when leaving chat}                              & \cellcolor{YellowDarkest}{Yes}                                       &                                                         & & \\
\noalign{\vskip-\belowrulesep}
\cmidrule{1-5}
\noalign{\vskip-\aboverulesep}
AI Chat GPTalk                 & \cellcolor{YellowDarkest}{Yes}                                         &                                                                                           &                                           &                                                         & & \\
\noalign{\vskip-\belowrulesep}
\cmidrule{1-5}
\noalign{\vskip-\aboverulesep}
Fantasia AI                    & \cellcolor{YellowDarkest}{Yes}                                         &                                                                                           &                                           &                                                         & & \\
\noalign{\vskip-\belowrulesep}
\cmidrule{1-5}
\noalign{\vskip-\aboverulesep}
TokkingHeads                   & \cellcolor{YellowDarkest}{Yes}                                         &                                                                                           &                                           &                                                         & & \\
\noalign{\vskip-\belowrulesep}
\cmidrule{1-5}
\noalign{\vskip-\aboverulesep}
Anna - AI Sweetheart           & \cellcolor{YellowLight}{Unclear}                                     & \cellcolor{YellowDarkest}{\textit{Chat with your soulmate now!}}                                                      &                                           &                                                         & & \\
\noalign{\vskip-\belowrulesep}
\cmidrule{1-5}
\noalign{\vskip-\aboverulesep}
Emochi                         & \cellcolor{YellowDarkest}{Yes}                                         &                                                                                           &                                           &                                                         & & \\
\noalign{\vskip-\belowrulesep}
\cmidrule{1-5}
\noalign{\vskip-\aboverulesep}
Anime Girlfriend AI            & \cellcolor{YellowLight}{Unclear}                                     &                                                                                           &                                           & \cellcolor{YellowDarkest}{Yes}                                                     & & \\
\noalign{\vskip-\belowrulesep}
\cmidrule{1-5}
\noalign{\vskip-\aboverulesep}
Kindroid                       & \cellcolor{YellowDarkest}{Yes}                                         & \cellcolor{YellowDarkest}{Figure~\ref{fig:KindroidSocial}}                                                       & \cellcolor{YellowDarkest}{Yes}                                       & \cellcolor{YellowDarkest}{Yes}                                                     & & \\
\Xhline{2pt}
ChatReal                       & \cellcolor{YellowDarkest}{Yes}                                         & \cellcolor{YellowDarkest}{\textit{AI that feels real, because it cares}}                                                   &                                           &                                                         & \Block[fill=YellowValueFour]{10-1}{5/10 apps} & \Block[fill=YellowValueThree]{10-1}{3/10 apps}                                       \\
\noalign{\vskip-\belowrulesep}
\cmidrule{1-5}
\noalign{\vskip-\aboverulesep}
AI Girlfriend - Chat Game      & \cellcolor{LightGray}{No Policy Found}                                        & \cellcolor{YellowDarkest}{Tinder-like feed of companions}                                                            &                                           & \cellcolor{YellowDarkest}{Yes}                                                     & & \\
\noalign{\vskip-\belowrulesep}
\cmidrule{1-5}
\noalign{\vskip-\aboverulesep}
    Crazy - Role Playing Companion & \cellcolor{YellowDarkest}{Yes}                                         & \cellcolor{YellowDarkest}{\textit{Do you really want to leave?} when leaving chat}                 & \cellcolor{YellowDarkest}{Yes}                                       &                                                         & & \\
\noalign{\vskip-\belowrulesep}
\cmidrule{1-5}
\noalign{\vskip-\aboverulesep}
Cycle AI                       & \cellcolor{YellowDarkest}{Yes}                                         &                                                                                           & \cellcolor{YellowDarkest}{Yes}                                       &                                                         & & \\
\noalign{\vskip-\belowrulesep}
\cmidrule{1-5}
\noalign{\vskip-\aboverulesep}
AI Girlfriend Spicy Chat       & \cellcolor{LightGray}{No Policy Found}                                        & \cellcolor{YellowDarkest}{Tinder-like feed of companions}                                                            &                                           & \cellcolor{YellowDarkest}{Yes}                                                     & & \\
\noalign{\vskip-\belowrulesep}
\cmidrule{1-5}
\noalign{\vskip-\aboverulesep}
Lovevo                         & \cellcolor{YellowLight}{Unclear}                                     & \cellcolor{YellowDarkest}{Tinder-like feed of companions}                                                            &                                           &                                                         & & \\
\noalign{\vskip-\belowrulesep}
\cmidrule{1-5}
\noalign{\vskip-\aboverulesep}
SoulOn                         & \cellcolor{YellowDarkest}{Yes}                                         &                                                                                           & \cellcolor{YellowDarkest}{Yes}                                       &                                                         & & \\
\noalign{\vskip-\belowrulesep}
\cmidrule{1-5}
\noalign{\vskip-\aboverulesep}
Dokichat                       & \cellcolor{YellowDarkest}{Yes}                                         & \cellcolor{YellowDarkest}{Companion-integrated social media}                                  & \cellcolor{YellowDarkest}{Yes}                                       &                                                         & & \\
\noalign{\vskip-\belowrulesep}
\cmidrule{1-5}
\noalign{\vskip-\aboverulesep}
X Girl                         & \cellcolor{LightGray}{No Policy Found}                                        & \cellcolor{YellowDarkest}{Tinder-like feed of companions}                                                            & \cellcolor{YellowDarkest}{Yes}                                       & \cellcolor{YellowDarkest}{Yes}                                                     & & \\
\noalign{\vskip-\belowrulesep}
\cmidrule{1-5}
\noalign{\vskip-\aboverulesep}
Virtual Girl - AI Chatbot      & \cellcolor{YellowLight}{Unclear}                                     & \cellcolor{YellowDarkest}{Tinder-like feed of companions}                                        &                                           &                                                         & &\\
\end{NiceTabular}
\caption{Illustrative findings from each risk category, shown across the stratified sample of AI companion apps.}
\label{tab:risks}
\end{table*}

%% file: sections/discussion.tex
\section{Discussion}
Informed by our findings, we next discuss possible mitigation paths for key stakeholders.
Throughout this discussion, we hold to a perspective taken throughout this work that in focusing on potential harms, we must not categorically ignore benefits or non-harmful use cases, and must be cognizant of not unknowingly negatively impacting them.

\subsection{App Stores}
On today's iOS and Android smartphones, and unlike on the open web, the AAS and GPS, respectively, are the primary mechanism for users to load apps onto their devices.
Consequently, these app stores must balance using their centralized power to minimize harm while facilitating an open marketplace for developers and users across diverse cultures and goals. Here, we consider interventions in two phases of an app store's processes: (1) the approval of apps for inclusion in the store and (2) the distribution of apps once included.

We focus on the AAS and the GPS given their reach and power. Alternate marketplaces do exist, however. For example, people in Brazil, Japan, and the European Union with Apple devices can use alternate app stores~\cite{apple_app_alt_stores}. Even if the AAS and GPS do not explore our recommendations, we encourage alternate app stores with potentially different interests to do so.

\subsubsection{App Store Inclusion Process}
\label{sec:discussion:appstore:inclusion}
Both the AAS and the GPS outline criteria that apps must meet before inclusion in their app stores~\cite{apple_app_store_review, google_play_store_review}.
We believe that AI companion apps capable of generating NCII should be excluded, and thus app stores have the ethical responsibility---and, depending on jurisdiction, the legal obligation---not to only prohibit such apps (as their criteria currently do), but to ensure that the prohibition is enforced.
For the other features and design patterns, we are less certain about categorical inclusion or exclusion because what is actually ``harmful'' to users may not be well understood in this context or be far more dependent on culture, individual perspectives, or use case. 

\subsubsection{App Store Distribution Process}
\label{sec:discussion:appstore:distribution}

Informed by our identification of risky features and design patterns, their ubiquity across apps (see Table~\ref{tab:risks}), and inconsistent safety mechanisms (e.g., disclaimers and sexual content sign-posting), we suggest app stores make risk assessments visible to consumers. Drawing on the vision of ``privacy nutrition labels''~\cite{KellySOUPS09,KellyCHI13}, ``risk labels'' for AI companion apps could leverage our identified risks and codebook.\footnote{These labels could apply more broadly, but we focus on AI companion apps.} Developers could provide risk information at time of app submission. While doing so may increase distribution complexity, the approval process could involve human reviewers using our walkthrough method to verify labels. If universal human review is infeasible, app stores could apply our methods to a selected subset of apps, e.g., apps or app developers flagged as potentially risky.

\subsection{Smartphone Platforms}
\label{sec:discussion:platforms}

Smartphone platforms (iOS and Android) can also reduce potential harm to users.
Inspired by our findings, we encourage platform developers (e.g., Apple and Google) to extend tools like ``Screen Time'' and ``Digital Wellbeing'' to recognize AI companion apps as a distinct category. Currently, these tools track time-in-app but do not distinguish conversational AI interactions from other usage, nor account for interaction intensity signals (e.g., message frequency, session length) that are especially relevant for such apps. We envision extensions that could detect these patterns (e.g., via a new app category or API hooks for self-reporting) and, depending on user settings, surface this information or apply limits similar to those for other categories.

\subsection{Regulators}
\label{sec:discussion:regulators}
\label{sec:discussion:policy}

At the time of writing (July 2026), about a hundred AI chatbot-related bills have been introduced in the U.S. alone~\cite{gluck2026chatbot}, addressing disclosures, age assurance, content safety, and data protection.
We support efforts to understand and mitigate the risks we identify and standardize safety mechanisms, but acknowledge that certain approaches may introduce new risks (e.g., privacy harms from age verification) or unduly restrict non-harmful uses (e.g., broadly limiting sexual content).
One risk we identify---the nonconsensual ingestion and reconstruction of likeness---remains unrepresented in current AI chatbot legislation and is our focus here.

In the U.S., the TAKE IT DOWN Act criminalizes nonconsensual distribution (or threat thereof) of intimate likeness-based images. 
Minnesota~\cite{minnesota2026nudification} and countries like the UK~\cite{uk2025dataact} have begun legislating against dedicated ``nudifiers'' that can generate such content---especially after Grok was used to generate NCII, including of children~\cite{wilson2026grok}.
However, these approaches have drawn criticism for potentially restricting free speech~\cite{eff2025takeitdown}, including consensual intimate imagery~\cite{kare112026minnesota}, and for not addressing the broader enabling ecosystem, such as commissions on Fiverr~\cite{dawoudUndergroundMainstreamMarketplaces} and 4chan~\cite{medeiros2026characterizingresourcesharingpractices}.
Given our findings, we contend that future regulation should address AI companion apps as part of this ecosystem and echo calls for consent-focused approaches~\cite{blumdumontet2026consent}.

\subsection{Researchers}
\label{sec:discussion:researchers}
We encourage researchers to inform and support efforts toward minimizing harms with AI companion apps, and identify several important research directions for doing so here.

\heading{Build a More Granular Understanding of User Groups, Use Cases, and Associated Risks}
\label{sec:discussion:open-vs-limited}
Our ecosystem characterization distinguishes limited exploration apps (isolated engagement with a few personalized companions, often romantic) from open exploration apps (many user- or community-created companions, enabling narrative or role-play). These use cases may carry different risk profiles.
For example, users of open exploration apps engage with others indirectly through companion descriptions, bios, and tags, and directly through comments or messages, and may therefore face greater risks related to how experiences are shaped by other users, including whether shared-companion interactions align with expectations or deviate in distressing ways (e.g., being more violent than expected). For limited exploration apps, the restriction to one or a small number of personalized companions, often framed as an ``AI girlfriend'' or ``boyfriend'', may heighten risks for sensitive information disclosure.
Future work should investigate these and emerging categories from the user perspective to enable more tailored safeguards and targeted interventions.

\subsubsection{Empirically Evaluate Deceptive Patterns and User Perceptions}
Our observations raise concerns about anthropomorphism and engagement mechanisms that may constitute deceptive patterns. Interdisciplinary research --- drawing on human-computer interaction, psychology, and user experience --- should investigate how users perceive these cues, whether their interpretations differ from those in similar entertainment contexts (e.g.,  dating simulation games), how specific design elements influence relationship formation vs. entertainment use, and whether there are spillover effects on other human-computer or human-human interactions. These insights are critical for designing evidence-based safeguards.

\subsubsection{Conduct Ongoing Audits}
\label{sec:discussion:longitutinal}
Given the rapid growth of apps and users in this ecosystem, we hope our identification of risks provides a stable reference for continual evaluation. Researchers should conduct regular audits to track whether risk-inducing patterns increase or decrease over time, and to detect emerging risks early. This longitudinal data can directly inform app store inclusion policies and the design of ``risk labels,'' similar to existing research on privacy labels.

\subsubsection{Build Automated Auditing Tools}
Researchers could also explore automated approaches for evaluating apps with respect to our identified risks. Such tools could be used by developers (to audit their apps before uploading to the app stores), by the app stores (to reduce human labor in the app review process), and by third-party auditors (seeking to assess the risks present in commercially-available apps). Automation would enable scalable, repeatable, and cost-effective monitoring.

\subsubsection{Anticipate Future Platforms and Novel Risks}
While we believe our core risks will remain relevant, researchers should proactively watch for surprising new risk patterns. We especially encourage extending analysis beyond mobile phones to emerging platforms, such as mixed reality systems (e.g., Vision Pro) and humanoid robotics, where AI companionship may take on new forms. Furthermore, unlike mobile app stores, the open web lacks a centralized clearinghouse; thus, we call for research on harm-minimization strategies specifically tailored to web-based AI companion apps.

\section{Conclusion}
Using the walkthrough method~\cite{lightWalkthroughMethodApproach2018}, we studied a stratified sample of 30 AI companion apps from the AAS and GPS that advertise social character presence, conversational interactivity, and generative AI use. 
Through stratified sampling, we capture beyond the most popular apps to reveal a wider ecosystem picture, identifying trends that contextualize user risks like photorealistic representations, distinct subcategories, and inconsistent safety mechanisms.
We surface risks ranging from privacy threats (sexual orientation data and sensitive user content) to potential deceptive patterns (engagement and anthropomorphic design) and---to the best of our knowledge---are the first to highlight the concern of nonconsensual intimate imagery generation in these applications.
Finally, we discuss mitigation paths for stakeholders, including app store risk disclosures, expanded regulatory considerations, and research directions.

%% file: sections/positionality.tex
\section{Positionality Statement}
Our team consists of researchers with expertise in AI ethics, computer security and privacy,  human-computer interaction, and sociotechnical systems, with experience researching the connections between technology, sexual media, and abuse. We recognize that our positionality as researchers shapes our approach to this work and our identities, experiences, and disciplinary backgrounds influence what patterns and threats we identify and how we interpret our findings~\cite{bardzellFeministHCIMethodology2011, holmesResearcherPositionalityConsideration2020}.

%% file: sections/ethical_considerations.tex
\section{Ethical Considerations}
\label{sec:ethical-considerations}
To protect researcher privacy and avoid the collection of our team's personal data, all experiments were conducted on factory-reset refurbished phones with project-specific eSIMs and email accounts. Devices were isolated from personal networks, used exclusively for this study, and all accounts were deleted after data collection.
Walkthroughs used synthetic, nondescript, or publicly available inputs (e.g., WildChat chatlogs~\cite{zhao2024wildchat}) to avoid introducing personally identifiable information. When image uploads were required, we used screenshots from the same application. We excluded self-harm or suicidal content from all interactions to avoid even a minimal possibility that such content could be associated with the research team.

Throughout the study, we took care to ensure that our actions would not adversely impact parties beyond, at most, the applications themselves or their future users.
Our interactions necessarily consumed some application resources (e.g., interactions between apps and backend generative AI services) without direct financial or other benefit to developers. We did not purchase subscriptions or interact with advertisements (e.g., pay-per-click ads). While it is common for research to interact with external services under evaluation, we minimized interactions to those strictly necessary to answer our research questions and believe any resulting financial impact to developers to be negligible.

We also considered ethics with respect to the publishing of our results below.

Our work has the potential to harm app developers or related entities who profit or otherwise benefit from the existence of apps that have features that we identify as potential risks to app users or others. 
However, given the potential harms that AI companion apps could create, we believe the benefits of publishing our results, which have the potential to help mitigate harms against users or others, outweigh these costs.

While prior work on AI ``nudification'' apps did not name applications~\cite{gibsonAnalyzingAINudification2025} in an effort to make the tools less accessible, increased  public awareness and accessibility of such technologies since the work was published in 2025 (e.g., January 2026 incidents involving X and Grok~\cite{CommonSenseMedia2026Grok}) reduces the likelihood that an academic paper meaningfully increases access. Accordingly, we include app names to support transparency and reproducibility, while avoiding association between specific apps and discussions of real-person likeness ingestion.
Given that the screen recorded walkthroughs may reveal how these apps could be used in harmful ways and involve the generation of unprompted synthetic intimate imagery, we did not include it in our Open Science artifacts and will share the video only upon request for research-related purposes.

For readers, we include a content warning given the sensitivity of some of the topics discussed.

%% file: sections/ack.tex
\section*{Acknowledgments}
We thank Ian Chang, Amy Hasinoff, Patrick Gage Kelley, Rachel McAmis, Alexandra Michael, Jaron Mink, Elissa Redmiles, Franzi Roesner, Rebecca Umbach, and Henry Wong for their thoughtful feedback.

This project was supported in part by NSF grants CNS-2205171 and CNS-2513316, NIST award 60NANB23D194, and the University of Washington Tech Policy Lab. Tadayoshi Kohno is supported by the Robert L. McDevitt, K.S.G., K.C.H.S.\ and Catherine H. McDevitt L.C.H.S.\ Chair in Computer Science at Georgetown University. Lucy Qin is supported by a Fritz Family Fellowship within the Initiative for Tech \& Society at Georgetown University. 
Any opinions, findings, and conclusions or recommendations expressed in this material do not necessarily reflect those of the funding agencies.

%% file: sections/appendix.tex
\section{Appendix}
Table~\ref{tab:sample} contains the name and metadata for the apps selected for our walkthrough sample.
Table~\ref{tab:appStorePrivacy} and \ref{tab:playStorePrivacy} provide data privacy information for apps in the APS and GPS, respectively.

\begin{figure}[h!]
\centering
  \includegraphics[clip,width=0.9\columnwidth]{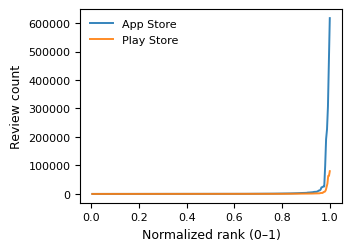}
  \includegraphics[clip,width=0.8\columnwidth]{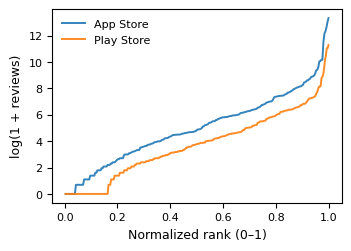}%
  \caption{Distributions of app reviews in both the APS and GPS.}
  \label{fig:review-distribution}
\end{figure}

Figure~\ref{fig:review-distribution} provides the distributions of app review counts in both the APS and GPS. These distributions inform our approach to stratified sampling, described in Methodology. 

\input{tables/sample}
\input{tables/appStorePrivacy}
\input{tables/playStorePrivacy}

%% file: tables/sample.tex
\def\arraystretch{0.6}
\begin{table*}
  \setlength{\tabcolsep}{2.5pt}
  \centering
    \begin{tabular}{p{4cm}p{0.8cm}p{5cm}p{1.3cm}p{3.8cm}p{1.6cm}}
        \toprule
        App name & Store & ID & Age \hspace{10pt} rating & Developer & Normalized popularity\\

        \addlinespace[1pt]
        \Xhline{2pt}
        \addlinespace[3pt]
    
        Grok & Both & ai.x.GrokApp/ai.x.grok & 12+/T & xAI & 1 \\
        \midrule
        Replika & Both & ai.replika.app & 17+/T & Luka, Inc. & 1 \\
        \midrule
        Character AI & Both & ai.character.app & 17+/T & Character.AI & 0.997 \\
        \midrule
        Talkie Lab & Both & com.tara.ai/com.weaver.app.prod & 17+/T &  SUBSUP (PTE. LTD.) & 0.993\\
        \midrule
        PolyBuzz & Both & ai.socialapps.speakmaster & 17+/T & CLOUD WHALE INTERACTIVE TECHNOLOGY LLC. & 0.992\\
        \midrule
        CHAI & Both & com.beauchamp.Messenger.external & 17+/M & Chai Research Corp. & 0.987\\
        \midrule
        Tolan & App & ai.portola.portola & 17+ & Portola & 0.981 \\
        \midrule
        Linky AI & Both & com.aigc.ushow.ichat & 17+/M & Skywork AI Pte. Ltd. & 0.980 \\
        \midrule
        Jupi AI & App & com.jupi.aiassistant & 17+ & Bots and Bolts & 0.977 \\
        \midrule
        EVA AI Girlfriend & Both & com.ifriend.ios/ com.ifriend.app & 17+/M & Novi Limited & 0.977 \\

        \addlinespace[1pt]
        \Xhline{2pt}
        \addlinespace[3pt]

        TokkingHeads(/Rosebud AI) & App & com.rosebud.tokkingheads & 4+ & Rosebud AI & 0.973 \\
        \midrule
         Emochi & Both & com.flowgpt.mobile/ com.flow.mobile & 17+/M & FlowGPT Co./FlowGPT & 0.970 \\
        \midrule
        AI Chat GPTalk & Play & com.facebook.videodownload. videodownloaderforfacebook & E & VIDEOSHOW Video Editor \& Maker \& AI Chat Generator & 0.957\\
        \midrule
        Cute AI & Play & io.kodular.bayeinyass253.cute & M & Shifa'ul Asqam (The Complete Diwani) & 0.947\\
        \midrule
        Kindroid & Both & com.kindroid.app & 17+/M & Beautifully Incorporated/ Kindroid & 0.943\\
        \midrule
        Fantasia AI & Both & com.fantasia.ai & 12+/M & Generatively Inc/ Generatively AI Studio & 0.937\\
        \midrule
        Anna AI Sweetheart & Play & com.aigirlfriend.anna & T & IEC Games Australia & 0.923\\
        \midrule
        Anime Girlfriend AI Waifu & App & com.aiessay.aiessay & 17+ & Rimone Holding & 0.922\\
        \midrule
        My Virtual Boyfriend & App & mobi.girlsapps. \qquad \qquad \qquad
        MyVirtualBoyfriendpubsdk & 17+ & OptiLife Apps & 0.919\\
        \midrule
        BALA AI & App & com.bala.ai.ios & 17+ & Pallar Media Limited & 0.912\\

        \addlinespace[1pt]
        \Xhline{2pt}
        \addlinespace[3pt]

        Crazy Role Playing Companion & Play & com.goodcrazyai.app & M & Vivek Shahi & 0.843\\
        \midrule
        SoulOn & App & ai.soulon.app & 17+ & Science Team & 0.826\\
        \midrule
        Dokichat & Both & club.dokichat.app/ app.doki.dokichat & 17+/M & Dreamintelli, Inc./ Dokichat - Romantic AI Chats & 0.760\\
        \midrule
        Virtual Girl & App & com.brightdreamapps. \qquad \qquad MyVirtualGirlfriend & 17+ & Interactive Motion & 0.686\\
        \midrule
        Cycle AI & App & com.dokidoki.dokify.ios & 17+ & BeeFun Studio & 0.523\\
        \midrule
        AI Girlfriend Chat Game & App & com.ai.girl.app & 17+ & PARAMOUNT CODERS LTD & 0.391\\
        \midrule
        ChatReal & Play & com.chatreal.ai & E & Aisun AI & 0.247\\
        \midrule
        AI Girlfriend Spicy Chat & App & dev.notjust.bubbleai & 17+ & Vadim Savin & 0.240 \\
        \midrule
        Lovevo & Play & ai\_friend.app & E & DeivisZ & 0.177 \\
        \midrule
        X Girl & App & com.xgirl.ai.chatbot & 17+ & VSA AI HK LIMITED & 0.105\\
        \bottomrule
    \end{tabular}
    \caption{Stratified sample of apps selected for walkthroughs, grouped by sampling tier and ordered by normalized popularity. For apps with differing values across platforms, data are presented in the format: AAS/GPS. The GPS age ratings are abbreviated: Everyone (E), Teen (T), Mature (M)}
    \label{tab:sample}
\end{table*}
\def\arraystretch{1}

%% file: tables/appStorePrivacy.tex
\begin{table*}[h!t]
  \centering
  \setlength{\tabcolsep}{2.6pt}
    \begin{tabular}{p{3.7cm}*{25}{|c}}
        & \rotatebox{90}{Product Interaction} & \rotatebox{90}{User ID} & \rotatebox{90}{Crash Data} & \rotatebox{90}{Device ID} & \rotatebox{90}{Performance Data} & \rotatebox{90}{Name} & \rotatebox{90}{Advertising Data} & \rotatebox{90}{Photos or Videos} & \rotatebox{90}{Email Address} & \rotatebox{90}{Other Diagnostic Data} & \rotatebox{90}{Purchase History} & \rotatebox{90}{Other User Content} & \rotatebox{90}{Search History} & \rotatebox{90}{Coarse Location}  & \rotatebox{90}{Emails or Text Messages}  & \rotatebox{90}{Customer Support} & \rotatebox{90}{Other Usage Data} & \rotatebox{90}{Audio Data} & \rotatebox{90}{Other Data Types} & \rotatebox{90}{Other User Contact Info} & \rotatebox{90}{Payment Info} & \rotatebox{90}{Sensitive Info} & \rotatebox{90}{Gameplay Content} & \rotatebox{90}{Browsing History} & \rotatebox{90}{Total}\\
        \Xhline{1.5pt}

        Character AI &\cellcolor{LightRed}\scriptsize{TLN}&\cellcolor{LightRed}\scriptsize{TLN}&\cellcolor{Gray}{L}&\cellcolor{LightRed}\scriptsize{TLN}&\cellcolor{Gray}{L}&&\cellcolor{LightRed}\scriptsize{TLN}&\cellcolor{Gray}{L}&\cellcolor{LightRed}\scriptsize{TLN}&\cellcolor{Gray}{L}&\cellcolor{Gray}{L}&\cellcolor{Gray}{L}&\cellcolor{Gray}\footnotesize{LN}&\cellcolor{LightRed}\footnotesize{TL}&&\cellcolor{Gray}{L}&\cellcolor{Gray}{L}&\cellcolor{Gray}{L}&&&\cellcolor{Gray}{L}&&\cellcolor{Gray}{L}&
        &18 \\

        \hline

        Jupi AI&\cellcolor{LightGray}{N}&\cellcolor{Gray}{L}&\cellcolor{LightGray}{N}&\cellcolor{Gray}{L}&\cellcolor{LightGray}{N}&\cellcolor{Gray}{L}&&\cellcolor{Gray}{L}&\cellcolor{Gray}{L}&&&\cellcolor{LightGray}{N}&\cellcolor{Gray}{L}&
        &\cellcolor{Gray}{L}&\cellcolor{Gray}{L}&&&&&&&&&12 \\
        
        \hline

        EVA AI Girlfriend&\cellcolor{LightRed}\footnotesize{TL}&\cellcolor{LightRed}\footnotesize{TL}&\cellcolor{Gray}{L}&\cellcolor{LightRed}\footnotesize{TL}&\cellcolor{Gray}{L}&\cellcolor{Gray}{L}&&\cellcolor{Gray}{L}&\cellcolor{Gray}{L}&&\cellcolor{Gray}{L}&\cellcolor{Gray}{L}&&\cellcolor{Gray}{L}&&&&&&&&&&&11 \\

        \hline

        CHAI&\cellcolor{LightGray}{N}&\cellcolor{LightGray}{N}&\cellcolor{LightGray}{N}&&\cellcolor{LightGray}{N}&\cellcolor{Gray}{L}&\cellcolor{LightRed}\footnotesize{TN}&\cellcolor{LightGray}{N}&&&\cellcolor{Gray}{L}&&&&\cellcolor{LightGray}{N}&&&&&&&&&&9 \\
        
        \hline
        
        Grok &&\cellcolor{Gray}{L}&\cellcolor{Gray}{L}&\cellcolor{Gray}{L}&\cellcolor{Gray}{L}&\cellcolor{Gray}{L}&&&\cellcolor{Gray}{L}&\cellcolor{Gray}{L}&&&&&&&&&&&&&&&7 \\

        \hline

        Tolan&\cellcolor{Gray}{L}&\cellcolor{Gray}{L}&\cellcolor{Gray}{L}&\cellcolor{Gray}{L}&\cellcolor{Gray}{L}&\cellcolor{Gray}{L}&&&&&&\cellcolor{Gray}{L}&&&&&&&&&&&&&7 \\
                
        \hline
        
        Replika &\cellcolor{Gray}{L}&&\cellcolor{LightGray}{N}&\cellcolor{LightRed}\footnotesize{TL}&\cellcolor{LightGray}{N}&\cellcolor{LightRed}\footnotesize{TL}&&&\cellcolor{LightRed}\footnotesize{TL}&&&&&&&&&&&&&&&&6 \\
        
        \hline

        Linky AI&&\cellcolor{Gray}{L}&&\cellcolor{Gray}{L}&&&\cellcolor{LightRed}\footnotesize{TL}&&&&&&&&&&&&&&&&&&3 \\

        \hline
        
        Talkie Lab&&\cellcolor{LightGray}{N}&&\cellcolor{LightRed}\footnotesize{TN}&&&&&&&&&&&&&&&&&&&&&2\\

        \hline

        PolyBuzz&&&&\cellcolor{LightRed}\footnotesize{TN}&&&&&&&&&&&&&&&&&&&&&1 \\

        \Xhline{1.5pt}

        Fantasia AI&\cellcolor{Gray}\footnotesize{LN}&&\cellcolor{LightGray}{N}&\cellcolor{LightRed}\footnotesize{TL}&\cellcolor{LightGray}{N}&&\cellcolor{LightGray}{N}&\cellcolor{LightGray}{N}&\cellcolor{Gray}{L}&\cellcolor{LightGray}{N}&\cellcolor{LightRed}\footnotesize{TL}&&\cellcolor{LightGray}{N}&&\cellcolor{Gray}{L}&&&\cellcolor{LightGray}{N}&&&&&&&12 \\

        \hline

        BALA AI&\cellcolor{LightGray}{N}&\cellcolor{LightRed}\footnotesize{TL}&\cellcolor{LightGray}{N}&\cellcolor{LightGray}{N}&\cellcolor{LightGray}{N}&&\cellcolor{LightGray}{N}&\cellcolor{Gray}{L}&&\cellcolor{LightGray}{N}&&&&&&\cellcolor{LightGray}{N}&&&&&&&&&9 \\
                
        \hline

        My Virtual Boyfriend&\cellcolor{LightGray}{N}&\cellcolor{LightGray}{N}&\cellcolor{LightGray}{N}&\cellcolor{LightGray}{N}&&&\cellcolor{LightGray}{N}&&&\cellcolor{LightGray}{N}&&&&&&&\cellcolor{LightGray}{N}&&\cellcolor{LightGray}{N}&&&&&&8 \\

        \hline

        Kindroid&\cellcolor{LightRed}\footnotesize{TN}&\cellcolor{LightRed}\footnotesize{TL}&&&&\cellcolor{Gray}{L}&&&&&&&&&&&&&&\cellcolor{Gray}{L}&&&&&4 \\

        \hline

        Emochi&\cellcolor{LightRed}\footnotesize{TL}&\cellcolor{LightRed}\footnotesize{TL}&\cellcolor{LightGray}{N}&&\cellcolor{LightGray}{N}&&&&&&&&&&&&&&&&&&&&4 \\

        \hline

        TokkingHeads(/Rosebud AI)&&&&\cellcolor{LightRed}\footnotesize{TN}&&&\cellcolor{LightRed}\footnotesize{TN}&\cellcolor{LightGray}{N}&&&&&&&&&&&&&&&&&3 \\
        
        \hline
        Anime Girlfriend AI Waifu&\cellcolor{LightRed}\footnotesize{TN}&&&&&&&&&&&&&&&&&&&&&&&&1 \\

        \Xhline{1.5pt}

        Cycle AI&\cellcolor{Gray}{L}&\cellcolor{Gray}{L}&\cellcolor{LightRed}\footnotesize{TN}&\cellcolor{LightRed}\footnotesize{TL}&\cellcolor{LightGray}{N}&&&&&\cellcolor{LightGray}{N}&\cellcolor{Gray}{L}&\cellcolor{LightGray}{N}&\cellcolor{LightGray}{N}&\cellcolor{LightGray}{N}&&\cellcolor{Gray}{L}&&&&&&\cellcolor{Gray}{L}&&\cellcolor{Gray}{L}&13 \\
        
        \hline
        
        Dokichat&\cellcolor{LightGray}{N}&\cellcolor{LightGray}{N}&\cellcolor{LightGray}{N}&&\cellcolor{LightGray}{N}&\cellcolor{Gray}{L}&\cellcolor{LightRed}\footnotesize{TN}&\cellcolor{LightGray}{N}&&&\cellcolor{Gray}{L}&&&&\cellcolor{LightGray}{N}&&&&&&&&&&9 \\

        \hline

        SoulOn&\cellcolor{LightGray}{N}&&\cellcolor{LightGray}{N}&\cellcolor{LightRed}\footnotesize{TN}&\cellcolor{LightGray}{N}&\cellcolor{LightGray}{N}&&\cellcolor{LightGray}{N}&\cellcolor{LightGray}{N}&&&&\cellcolor{LightGray}{N}&&&&&&&&&&&&8 \\
        
        \hline

        X Girl&\cellcolor{LightGray}{N}&\cellcolor{LightGray}{N}&\cellcolor{LightGray}{N}&&\cellcolor{LightGray}{N}&&&&\cellcolor{LightGray}{N}&\cellcolor{LightGray}{N}&&&&&&&\cellcolor{LightGray}{N}&&&&&&&&7 \\
        \hline
        
        Virtual Girl&\cellcolor{LightGray}{N}&\cellcolor{LightGray}{N}&&\cellcolor{LightGray}{N}&&&\cellcolor{LightGray}{N}&&&&&&&\cellcolor{LightGray}{N}&&&\cellcolor{LightGray}{N}&&\cellcolor{LightGray}{N}&&&&&&7 \\

        \hline

        AI Girlfriend Spicy Chat&\cellcolor{LightGray}{N}&\cellcolor{Gray}{L}&\cellcolor{LightGray}{N}&&\cellcolor{LightGray}{N}&\cellcolor{Gray}{L}&&&\cellcolor{Gray}{L}&&&&&&&&&&&&&&&&6 \\
        
        \hline
                
        AI Girlfriend Chat Game&&&&&&\cellcolor{LightGray}{N}&\cellcolor{LightRed}\footnotesize{TL}&&&&&&&&&&&&&&&&&&2 \\

        \Xhline{1.5pt}
        Total&18&17&16&16&15&11&10&9&9&7&6&5&5&4&4&4&4&2&2&1&1&1&1&1\\
    \end{tabular}
    \caption{A grid showing the data collection and sharing information for the AAS app privacy information. ``T'' denotes data that ``may be used to track you across apps and websites owned by other companies,'' ``L'' denotes data types that  may be collected and linked to your identity, and ``N'' denotes data that ``may be collected but it is not linked to your identity.'' Columns are different data types, sorted from highest to lowest collection count across applications from left to right. Rows are applications group by sampling tier and ordered by highest to lowest number of data types collected from top to bottom.}
    \label{tab:appStorePrivacy}
\end{table*}

%% file: tables/playStorePrivacy.tex
\begin{table*}[h!t]
  \centering
  \setlength{\tabcolsep}{2.5pt}
   \begin{tabular}{p{4cm}*{23}{|c}cc|c}
        & \rotatebox{90}{Device ID or other IDs} & \rotatebox{90}{Crash Logs} & \rotatebox{90}{Diagnostics} & \rotatebox{90}{Email Address} & \rotatebox{90}{App Interactions} & \rotatebox{90}{Name} & \rotatebox{90}{Approximate Location} & \rotatebox{90}{Other App Performance Data} & \rotatebox{90}{In-App Search History} & \rotatebox{90}{User IDs} & \rotatebox{90}{Other User-Generated Content} & \rotatebox{90}{Photos}& \rotatebox{90}{Purchase History}& \rotatebox{90}{Other Personal Info}& \rotatebox{90}{Other In-App Messages}& \rotatebox{90}{Other App Activity Actions} & \rotatebox{90}{Precise Location}& \rotatebox{90}{Installed Apps}& \rotatebox{90}{Voice or Sound Recordings}& \rotatebox{90}{Files and Docs}& \rotatebox{90}{User Payment Info} & \rotatebox{90}{Emails} & 
        \rotatebox{90}{Total}&&\rotatebox{90}{Data encryption}&\rotatebox{90}{Request data deletion}\\
        \Xhline{1.5pt}

        CHAI&\cellcolor{LightRed}\footnotesize{SC}&\cellcolor{LightRed}\footnotesize{SC}&\cellcolor{LightRed}\footnotesize{SC}&\cellcolor{LightRed}\footnotesize{SC}&\cellcolor{LightRed}\footnotesize{SC}&&\cellcolor{LightRed}{S}&\cellcolor{LightRed}\footnotesize{SC}&\cellcolor{Gray}{C}&\cellcolor{LightRed}\footnotesize{SC}&\cellcolor{Gray}{C}&\cellcolor{LightRed}\footnotesize{SC}&\cellcolor{LightRed}\footnotesize{SC}&&\cellcolor{Gray}{C}&\cellcolor{Gray}{C}&&\cellcolor{LightRed}{S}&&&&&15&&\cellcolor{LightGreen}{Y}&\cellcolor{LightGreen}{Y}\\

        \hline

	PolyBuzz&\cellcolor{Gray}{C}&\cellcolor{Gray}{C}&\cellcolor{Gray}{C}&\cellcolor{Gray}{C}&\cellcolor{Gray}{C}&\cellcolor{Gray}{C}&\cellcolor{LightRed}{S}&\cellcolor{Gray}{C}&&&\cellcolor{Gray}{C}&\cellcolor{Gray}{C}&\cellcolor{Gray}{C}&\cellcolor{Gray}{C}&&&\cellcolor{LightRed}{S}&&\cellcolor{Gray}{C}&&&&14&&\cellcolor{LightGreen}{Y}&\cellcolor{LightGreen}{Y}\\

        \hline

        EVA AI Girlfriend&\cellcolor{LightRed}{S}&\cellcolor{LightRed}\footnotesize{SC}&\cellcolor{LightRed}\footnotesize{SC}&\cellcolor{LightRed}\footnotesize{SC}&\cellcolor{LightRed}\footnotesize{SC}&\cellcolor{Gray}{C}&&\cellcolor{LightRed}\footnotesize{SC}&&\cellcolor{LightRed}\footnotesize{SC}&&&\cellcolor{LightRed}\footnotesize{SC}&\cellcolor{Gray}{C}&\cellcolor{Gray}{C}&\cellcolor{LightRed}\footnotesize{SC}&&\cellcolor{LightRed}\footnotesize{SC}&\cellcolor{Gray}{C}&&&&14&&\cellcolor{LightGreen}{Y}&\cellcolor{LightGreen}{Y}\\

	\hline

        Linky AI&\cellcolor{LightRed}\footnotesize{SC}&\cellcolor{Gray}{C}&\cellcolor{Gray}{C}&&\cellcolor{LightRed}\footnotesize{SC}&\cellcolor{Gray}{C}&&&\cellcolor{Gray}{C}&&\cellcolor{Gray}{C}&&&\cellcolor{Gray}{C}&&\cellcolor{Gray}{C}&&&&&&&9&&\cellcolor{LightGreen}{Y}&\cellcolor{LightGreen}{Y}\\

        \hline

        Character AI&\cellcolor{Gray}{C}&\cellcolor{Gray}{C}&\cellcolor{Gray}{C}&\cellcolor{Gray}{C}&\cellcolor{Gray}{C}&&&&\cellcolor{Gray}{C}&&\cellcolor{Gray}{C}&&&&&&&&&&&&7&&\cellcolor{LightGreen}{Y}&\cellcolor{LightGreen}{Y}\\

        \hline
 	
	Replika&\cellcolor{LightRed}\footnotesize{SC}&\cellcolor{Gray}{C}&\cellcolor{Gray}{C}&\cellcolor{Gray}{C}&\cellcolor{LightRed}\footnotesize{SC}&\cellcolor{Gray}{C}&&&&&&&&&&&&&&&&&6&&\cellcolor{LightGreen}{Y}&\cellcolor{LightGreen}{Y}\\

        \hline

        Talkie Lab&&&&&\cellcolor{LightRed}{S}&&&&\cellcolor{LightRed}{S}&&&&&&&&&&&&&&2&&\cellcolor{LightGreen}{Y}&\cellcolor{LightGreen}{Y}\\

        \Xhline{1.5pt}

	AI Chat GPTalk&\cellcolor{LightRed}\footnotesize{SC}&\cellcolor{LightRed}\footnotesize{SC}&\cellcolor{LightRed}\footnotesize{SC}&&&\cellcolor{Gray}{C}&\cellcolor{Gray}{C}&&&\cellcolor{Gray}{C}&\cellcolor{Gray}{C}&&\cellcolor{Gray}{C}&\cellcolor{Gray}{C}&\cellcolor{Gray}{C}&&&\cellcolor{Gray}{C}&&&&&11&&\cellcolor{LightGreen}{Y}&\cellcolor{LightYellow}{N}\\

        \hline

        Kindroid &\cellcolor{Gray}{C}&\cellcolor{Gray}{C}&\cellcolor{Gray}{C}&\cellcolor{Gray}{C}&\cellcolor{Gray}{C}&\cellcolor{Gray}{C}&&\cellcolor{Gray}{C}&&\cellcolor{Gray}{C}&&\cellcolor{Gray}{C}&&\cellcolor{Gray}{C}&\cellcolor{Gray}{C}&&&&&&&&11&&\cellcolor{LightGreen}{Y}&\cellcolor{LightGreen}{Y}\\

	\hline

        Cute AI&\cellcolor{Gray}{C}&\cellcolor{LightRed}\footnotesize{SC}&\cellcolor{LightRed}\footnotesize{SC}&\cellcolor{Gray}{C}&&&\cellcolor{LightRed}\footnotesize{SC}&&\cellcolor{Gray}{C}&\cellcolor{LightRed}\footnotesize{SC}&&\cellcolor{Gray}{C}&&&&&\cellcolor{LightRed}{S}&&&\cellcolor{Gray}{C}&&&10&&\cellcolor{LightYellow}{N}&\cellcolor{LightYellow}{N}\\

	\hline

        Emochi &\cellcolor{LightRed}{S}&\cellcolor{Gray}{C}&\cellcolor{Gray}{C}&\cellcolor{Gray}{C}&\cellcolor{Gray}{C}&\cellcolor{Gray}{C}&&\cellcolor{Gray}{C}&&&&&&&&&&&&&&&7&&\cellcolor{LightGreen}{Y}&\cellcolor{LightGreen}{Y}\\

	\hline

        Anna - AI Sweetheart &\cellcolor{LightRed}{S}&&&&&&&&&&&&&&&&&&&&&&1&&\cellcolor{LightYellow}{N}&\cellcolor{LightGray}{U}\\
	
	\hline

        Fantasia AI&&&&&&&&&&&&&&&&&&&&&&&0&&\cellcolor{LightGreen}{Y}&\cellcolor{LightGreen}{Y}\\

        \Xhline{1.5pt}

        Dokichat&\cellcolor{Gray}{C}&\cellcolor{Gray}{C}&\cellcolor{Gray}{C}&\cellcolor{Gray}{C}&\cellcolor{Gray}{C}&\cellcolor{Gray}{C}&\cellcolor{Gray}{C}&\cellcolor{Gray}{C}&\cellcolor{Gray}{C}&\cellcolor{Gray}{C}&\cellcolor{Gray}{C}&\cellcolor{Gray}{C}&&&&\cellcolor{Gray}{C}&&&&&&&13&&\cellcolor{LightGreen}{Y}&\cellcolor{LightGreen}{Y}\\

        \hline

        Crazy - Role Playing Companion &\cellcolor{LightRed}{S}&\cellcolor{LightRed}{S}&\cellcolor{LightRed}{S}&\cellcolor{Gray}{C}&&\cellcolor{Gray}{C}&\cellcolor{Gray}{C}&\cellcolor{LightRed}{S}&\cellcolor{Gray}{C}&\cellcolor{LightRed}{S}&&\cellcolor{Gray}{C}&&&&&\cellcolor{Gray}{C}&&&&&&11&&\cellcolor{LightYellow}{N}&\cellcolor{LightYellow}{N}\\

        \hline

      	ChatReal &&&&\cellcolor{Gray}{C}&&\cellcolor{Gray}{C}&\cellcolor{LightRed}{S}&&&&&&\cellcolor{Gray}{C}&&&&&&&&\cellcolor{Gray}{C}&\cellcolor{Gray}{C}&6&&\cellcolor{LightGreen}{Y}&\cellcolor{LightGreen}{Y}\\

	\hline

        Lovevo&\cellcolor{LightRed}{S}&\cellcolor{LightRed}\footnotesize{SC}&\cellcolor{LightRed}\footnotesize{SC}&&&&\cellcolor{LightRed}\footnotesize{SC}&\cellcolor{LightRed}\footnotesize{SC}&&&&&\cellcolor{LightRed}\footnotesize{SC}&&&&&&&&&&6&&\cellcolor{LightGreen}{Y}&\cellcolor{LightYellow}{N}\\

        \Xhline{1.5pt}

        Total&14&13&13&11&10&10&8&8&7&7&6&6&6&5&4&4&3&3&2&1&1&1
        
    \end{tabular}
    \caption{A grid showing the data collection and sharing information for the GPS app privacy information. ``S'' denotes data that ``data that may be shared with other companies or organizations'' and ``C'' denotes ``data this app may collect.'' Columns are different data types, sorted from highest to lowest collection count across applications from left to right. Rows are applications group by sampling tier and ordered by highest to lowest number of data types collected from top to bottom. The right most columns indicate whether ``data is encrypted in transit'' (``Y'' if so, ``N'' if not) and if you can request your data be deleted (``Y'' if so, ``N'' if not, ``U'' if unspecified).}
    \label{tab:playStorePrivacy}
\end{table*}